\documentclass[letterpaper,twocolumn,english,showpacs,pra,footinbib,superscriptaddress]{revtex4-1}
\usepackage[latin9]{inputenc}
\usepackage{color}
\usepackage{babel}
\usepackage{float}
\usepackage{amsmath}
\usepackage{amsthm}
\usepackage{amssymb}
\usepackage{stmaryrd}
\usepackage{graphicx}
\usepackage[unicode=true,
 bookmarks=false,
 breaklinks=true,pdfborder={0 0 1},backref=section,colorlinks=true]
 {hyperref}
\hypersetup{
 colorlinks,plainpages=true,hypertexnames=false,pageanchor=true,linkcolor={blue},citecolor={blue},urlcolor={blue},anchorcolor={black}}

\makeatletter

\usepackage{babel}

\usepackage{epsfig}\usepackage{dcolumn}\usepackage{mathrsfs}\usepackage{pifont}\usepackage{amsthm}\usepackage{bm}\usepackage{latexsym}\usepackage{amsfonts}

\newcommand{\beq}{\begin{equation}}
\newcommand{\eeq}{\end{equation}}
\newcommand{\beqa}{\begin{eqnarray}}
\newcommand{\eeqa}{\end{eqnarray}}

\makeatother

\begin{document}
\title{Noise assisted quantum coherence protection in hierarchical environment}
\author{Xinyu Zhao}
\affiliation{Fujian Key Laboratory of Quantum Information and Quantum Optics (Fuzhou
University), Fuzhou 350116, China}
\affiliation{Department of Physics, Fuzhou University, Fuzhou 350116, China}
\author{Yong-hong Ma}
\email{myh\underline{ }dlut@126.com}

\affiliation{School of Science, Inner Mongolia University of Science and Technology,
Baotou 014010, People's Republic of China}
\author{Yan Xia}
\email{xia-208@163.com}

\affiliation{Fujian Key Laboratory of Quantum Information and Quantum Optics (Fuzhou
University), Fuzhou 350116, China}
\affiliation{Department of Physics, Fuzhou University, Fuzhou 350116, China}
\begin{abstract}
In this paper, we investigate coherence protection of a quantum system
coupled to a hierarchical environment by utilizing noise. As an example,
we solve the Jaynes-Cummings (J-C) model in presence of both a classical
and a quantized noise. The master equation is derived beyond the Markov
approximation, where the influence of memory effects from both noises
is taken into account. More importantly, we find that the performance
of the coherence protection sensitively depends on the non-Markovian
properties of both noises. By analyzing the mathematical mechanism
of the coherence protection, we show the decoherence caused by a non-Markovian
noise with longer memory time can be suppressed by another Markovian
noise with shorter memory time. Last but not least, as an outlook,
we try to analyze the connection between the atom-cavity entanglement
and the atomic coherence, then discuss the possible clue to search
the required noise. The results presented in this paper show the possibility
of protecting coherence by utilizing noise and may open a new path
to design noise-assisted coherence protection schemes.
\end{abstract}
\maketitle

\section{\label{sec:Intorduction}Introduction}

Quantum coherence is a unique feature that makes the quantum realm
to be different from the classical world \cite{Zurek2003RMP}. It
is also a precious resource in quantum computation. Particularly,
more research has focused on quantum coherence since the quantum computational
advantage has been observed in experiments \cite{Arute2019Nature,Zhong2020Science}.
However, quantum coherence is fragile when the quantum system inevitably
interacts with its environment \cite{Zurek1991PT,Yu2004PRL,Yu2009Science}.
To protect quantum coherence, various smart schemes are proposed to
eliminate the decoherence caused by noises. For example, quantum error
correction codes \cite{Shor1995PRA,Steane1996PRL,Bennett1996PRA,Webster2020QV,Fowler2012PRA}
is a direct analog to the classical error correction, where auxiliary
qubits are employed as redundancy. Besides, one can also cancel the
decoherence by inserting control pulses periodically, which dates
back to the spin-echo technique and extends to a family of dynamical
decoupling schemes \cite{Hahn1950PR,Uhrig2007PRL,Viola1998PRA,Viola1999PRL,Khodjasteh2005PRL}.
Another widely studied scheme is the quantum feedback control \cite{Wiseman1993PRL,Wiseman1994PRA,Zhang2017PR,Geremia2004Science},
in which one can monitor the decoherence process and use feedback
operations to compensate the loss of coherence. Certainly, there are
several other ways to fight against noise, such as decoherence free
space, quantum weak measurement reversal \cite{Koashi1999PRL,Korotkov2006PRL,Korotkov2010PRA,Ashhab2010PRA},
environmental-assisted error correction \cite{Nagali2009IJQI,Trendelkamp-Schroer2011PRA,Gregoratti2003JMO,Wang2014PRA},
etc. However, all these methods have their own limitations and the
implementation of these schemes requires extra physical resources.
Particularly, when multiple noises appear, more resources (more redundancy
qubits, multiple layers of control pulses, extra feedback loops, or
a larger decoherence-free space, etc.) are required to eliminate all
noises.

In contrast to active coherence protection schemes discussed above,
it is also valuable to investigate the properties of noise and its
impact on decoherence \cite{Clerk2010RoMP}. Particularly, in the
presence of multiple noises, it is interesting to ask whether the
decoherence caused by a noise can be eliminated by another noise.
This may imply an alternative way to weaken the decoherence of quantum
system purely by utilizing the properties of noises \cite{Corn2009QIP,jing2013SR},
which is certainly beneficial for the potential error correction operations
in the next step. A few successful examples have been shown in Refs.
\cite{jing2013SR,Jing2014PRA,Jing2018PRA,Khodorkovsky2008PRL}, although
the research in this direction is still limited. Two questions have
to be answered. (i) In what configuration, adding a noise can weaken
the decoherence caused by another noise, namely for a given noise,
how to add a second noise to reduce the decoherence. (ii) What are
the impacts of the properties (e.g., the memory time) of the two noises.
Particularly, whether the non-Markovian properties are helpful to
reduce the decoherence.

\begin{figure}
\noindent \begin{centering}
\includegraphics[width=1\columnwidth]{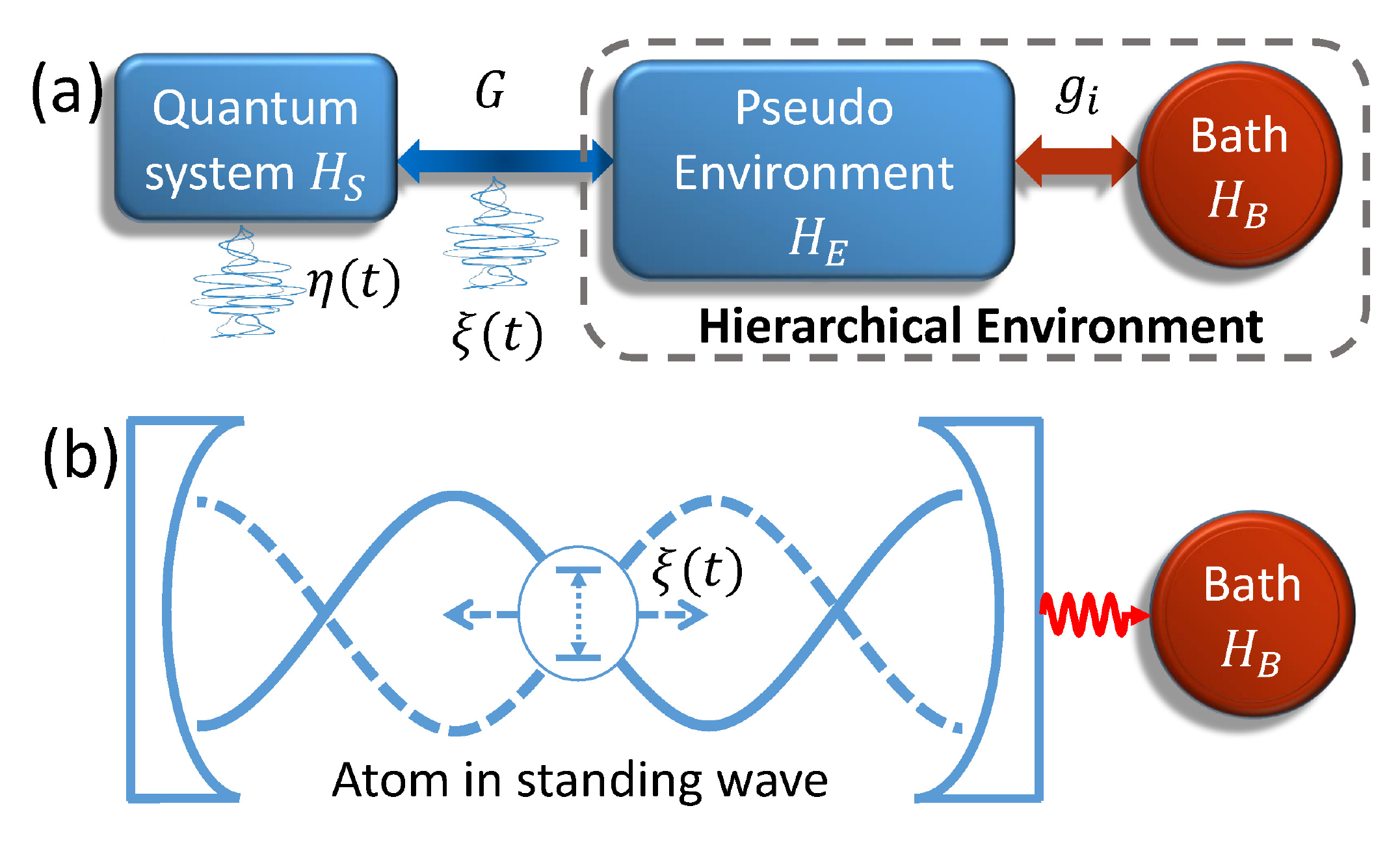} 
\par\end{centering}
\caption{(a) Schematic diagram of a quantum system coupled to a hierarchical
environment. (b) An example of pseudo-environment: J-C model coupled
to an external bath. The atom can be regarded as the quantum system
and the cavity plays the role of pseudo-environment. The coupling
strength between atom and cavity (pseudo-environment) depends on the
position of the atom in the cavity \cite{Natali2007PRA,Wu1997PRL,Mabuchi1999APB,Duan2003PRA,Hood2000Science}.}

\label{fig:1} 
\end{figure}

In this paper, we try to find some clues to answer these two questions
by analyzing the decoherence of a particular physical model, a quantum
system coupled to a hierarchical environment as shown in Fig.~\ref{fig:1}~(a).
Besides the quantized noise from the bath $H_{{\rm B}}$, a second
noise $\xi(t)$ or $\eta(t)$ is introduced to reduce the decoherence.
First of all, we derive the master equation beyond the Markov approximation
in presence of both noises, which ensures an accurate evolution taking
non-Markovian memory effects into account. The derivation of time-convolutionless
master equation (with both classical and quantized noises) itself
is valuable in theoretical studies. This may provide a systematic
way to derive master equations and study the mutual (indirect) interactions
of two noises in the future. Second, as the central results of the
paper, we show the decoherence sensitively depends on the properties
of both noises (one is from the quantized bosonic bath, the other
is the classical noise). In particular, the non-Markovian properties
like the memory time of the two noises directly determine that adding
a second noise leads to a positive or negative effect of coherence
protection. This partially answers the question (ii) and emphasizes
the importance of non-Markovian behaviors \cite{Zhao2011PRA,Zhao2019OEa,Mu2016PRA}.
Third, we also analyze the performance of the coherence protection
for different noises, $\xi(t)$ or $\eta(t)$. In a brief summary,
the mechanism of the coherence protection can be mathematically interpreted
as the contribution of a slow-varying function can be eliminated by
a fast-varying function in the time-integral. Therefore, it may imply
the possibility of using a high-frequency noise to suppress a low-frequency
noise. Last but not least, as an outlook, we briefly discuss the relation
between decoherence and the entanglement generation between the system
and the ``pseudo-environment''. This may provide useful clues to
answer the question (i). At least, it provides a possible direction
to search what type of a second noise can protect the coherence.

The rest of the paper is organized as follows: In Sec.~\ref{sec:2Model},
we illustrate our coherence protection scheme in an example model
and solve the model beyond the Markov approximation. Non-Markovian
master equations are derived in presence of both classical and quantized
noises. In Sec.~\ref{sec:3}, we show the coherence can be protected
by adding another noise and analyze how the properties of the two
noises affect the performance of the coherence protection. In Sec.~\ref{sec:5},
we draw a conclusion and discuss several valuable research topics
in the future.

\noindent 

\section{\label{sec:2Model}Model and solution}

\subsection{The model: hierarchical environment}

In order to illustrate the feasibility of using noise to suppress
decoherence caused by another noise, we focus on a common scenario
that a quantum system is coupled to a hierarchical environment as
shown in Fig.~\ref{fig:1}~(a). Similar to Ref. \cite{Qiao2019SCPM&A},
an artificial quantum system $H_{{\rm E}}$ is introduced as a ``pseudo-environment''
to connect the system $H_{{\rm S}}$ and the bath $H_{{\rm B}}$.
The reason to consider such a ``pseudo-environment'' is that the
coupling between $H_{{\rm S}}$ and $H_{{\rm E}}$ is typically easier
to be manipulated (e.g., adding a noise) than the direct coupling
to $H_{{\rm B}}$. In this scenario, $H_{{\rm E}}+H_{{\rm int}}+H_{{\rm B}}$
is regarded as a hierarchical environment of $H_{{\rm S}}$. The randomness
of the bath can be regarded as a quantum noise, resulting a coherence
loss when taking the trace of environmental degrees of freedom \cite{Zurek2003RMP,Zurek1991PT,Gardiner2004Book}.
Besides, we also consider a second noise originates from the fluctuation
of the classical fields. Here, we focus on two types. On is noise
$\xi(t)$ in the coupling $G$ between $H_{{\rm S}}$ and $H_{{\rm E}}$
and the other is noise $\eta(t)$ directely coupled to the system.

For different physical system the dominant noise could be very different.
In the following sections, we will discuss the impacts of both types
of noises. In Sec.\ref{sec:3}, we mainly focus on the noise $\xi(t)$
with a detailed discussion on the influence of its non-Markovian properties.
In Sec.\ref{sec:eta}, we make a brief discussion on the noise $\eta(t)$
and its difference from $\xi(t)$.

The hierarchical environment configuration in Fig.~\ref{fig:1}~(a)
is very common in many physical systems. One example in cavity-QED
\cite{Rempe1987PRL,Eberly1980PRL,Hood2000Science,Natali2007PRA,Zheng2000PRL}
system is given in Fig.~\ref{fig:1}~(b). It can be described by
the Jaynes-Cummings (J-C) model \cite{Jaynes1963IEEE} coupled to
an external environment. The Hamiltonian can be written as 
\begin{equation}
H_{\mathrm{tot}}=H_{{\rm JC}}+H_{\mathrm{B}}+H_{\mathrm{int}},\label{eq:Htot}
\end{equation}
where 
\begin{equation}
H_{{\rm JC}}=\frac{\omega(t)}{2}\sigma_{z}+\Omega a^{\dagger}a+G(t)a\sigma_{+}+G(t)a^{\dagger}\sigma_{-},\label{eq:HJC}
\end{equation}
\begin{equation}
H_{\mathrm{B}}=\sum_{i}\omega_{i}b^{\dagger}b_{i},\label{eq:HB}
\end{equation}
\begin{equation}
H_{\mathrm{int}}=\sum_{i}g_{i}(ab_{i}^{\dagger}+a^{\dagger}b_{i}).\label{eq:Hint}
\end{equation}
In Eq.~(\ref{eq:HJC}), $H_{{\rm JC}}$ is the J-C Hamiltonian \cite{Shore1993JMO,Jaynes1963IEEE,Rempe1987PRL}
describing the interaction between a two-level atom and a cavity,
where $a$ is the annihilation operator of the cavity mode with frequency
$\Omega$, $\sigma_{z}=|e\rangle\langle e|-|g\rangle\langle g|$,
$\sigma_{+}=|e\rangle\langle g|$, and $\sigma_{-}=|g\rangle\langle e|$
are the atomic operators. The cavity leakage is described by an interaction
$H_{\mathrm{int}}$ in Eq.~(\ref{eq:Hint}) with the external bosonic
bath $H_{\mathrm{B}}$ in Eq.~(\ref{eq:HB}), where $b_{i}$ are
the annihilation operators of different modes in the bath. The cavity
plus the external bath can be regarded as a hierarchical environment,
and the cavity plays a role of pseudo-environment ($H_{{\rm S}}=\frac{\omega}{2}\sigma_{z}$,
$H_{{\rm E}}=\Omega a^{\dagger}a$).

Besides the cavity-QED example shown in Fig.~\ref{fig:1}~(b), there
may be other physical realizations of the Hamiltonian (\ref{eq:HJC}).
For example, in the circuit-QED system \cite{Devoret2013Science},
such a Hamiltonian can be used to describe the interaction between
artificial atoms \cite{You2005PT,You2011Nature,Kastner1993PT} and
quantum harmonic oscillators. Besides, when the semiconductor quantum
dots are coupled to a cavity in recent the experimental progress \cite{Burkard2020NRP},
the system can be also described by the J-C model \cite{Burkard2020NRP}.
A detailed discussion on alternative physical system of J-C models
is given in Appendix~\ref{app:DQD}.

It is also worth to note that the J-C model is not the only example
of a hierarchical environment. In solid state quantum dots \cite{Hanson2007RMP},
the electron spin degree of freedom is naturally coupled to the outside
environment through the spin-orbit interaction \cite{Bychkov1984JPC,Dresselhaus1955PR},
where the orbital degree of freedom naturally plays the role of the
pseudo-environment. One particular example is also discussed in Appendix~\ref{app:DQD}. 

The key point of achieving coherence protection by using noise is
the time dependent coupling $G(t)$ and frequency $\omega(t)$, however,
in different physical system, the realizations of the noises are also
different. For example, in the cavity-QED system shown in Fig.~\ref{fig:1}~(b),
the noise in $G(t)$ may originates from the motion of the atom in
the cavity. If the atom is near the anti-node or the node of the standing
wave in the cavity, the coupling strength may be quite different \cite{Natali2007PRA,Wu1997PRL,Mabuchi1999APB,Duan2003PRA,Hood2000Science}.
Only considering the $x$-component freedom, it can be written as
\cite{Wu1997PRL,Mabuchi1999APB} 
\begin{equation}
G(t)=G_{0}\sin\left\{ k\left[x_{0}+\xi(t)\right]\right\} ,\label{Gt}
\end{equation}
where $k$ is the wave number of the standing wave and $x(t)=x_{0}+\xi(t)$
is the position of the atom in $x$-direction. Suppose the atom moves
in the cavity randomly, the function $x(t)$ can be described by a
stochastic process (e.g., vibration or Brownian motion of the atom).
In contrast, in the circuit-QED system discussed in Appendix~\ref{app:DQD},
The tunable (time-dependent) coupling can be realized by electrical
signals (through external flux of the coupler) \cite{Tian2008NJP}.
Some other tunable coupling schemes such like flux qubit coupled to
a resonator are also discussed in Refs. \cite{Brink2005NJP,Averin2003PRL,Plourde2004PRB,Hime2006Science}.
Therefore, the noise introduced in the coupling can be either noise
in natural world like the Brownian motion of an atom or some artificial
noises like random electrical pulses.

The randomness in $\omega(t)$ is widely studied in circuit-QED system
\cite{Kjaergaard2020ARoCMP,You2005PT,You2011Nature}, and it often
originates from the charge noise. In the example discussed in Appendix~\ref{app:DQD},
the fluctuation of magnetic field can also cause time-dependent $\omega(t)$
\cite{Burkard2020NRP,Hanson2007RMP}. Here, we assume
\begin{equation}
\omega(t)=\omega_{0}+\eta(t),\label{eq:wt}
\end{equation}
where the stochastic function $\eta(t)$ represents the noise.

In this paper, we only assume there is one type of classical noise,
either $\xi(t)$ or $\eta(t)$ is applied. By following the same procedure
in Appendix~\ref{app:MEQ}, one can certainly solve the case both
$\xi(t)$ and $\eta(t)$ are non-zero, but it is left for the future
studies.

\subsection{Solution}

Despite the coherence protection we will discuss in the following
sections, the solution of this model itself is also valuable since
the model attracts so much research interests in theoretical and experimental
studies \cite{OConnell2010Nature,Mi2018N,Chen2021PRL,Chen2017PRA,Chen2020arXiv,Zhao2020PRA,Xiong2019OE,Chen2017OE}.
For instance, a better understanding of this model may contribute
to solving the non-Markovian measurement problem which can be applied
in gravitational wave detection \cite{Chen2013JPB,Yang2012PRA}. However,
previous studies \cite{Breuer2007Book} are often based on Markov
approximation. In this paper, we employ the ``non-Markovian quantum
state diffusion'' (NMQSD) approach to derive a fundamental dynamic
equation of the system. Then, we obtain the master equation by taking
the statistical mean over all noises. The master equation simultaneously
contains the impacts from two noises, $\xi(t)$ {[}or $\eta(t)${]}
plus the noise from $H_{{\rm B}}$. It provides a powerful tool to
investigate the non-Markovian dynamics under the influences of both
the classical noise and the quantized noise. 

By expanding the environmental degrees of freedom with the multi-mode
Bargemann state $|z\rangle\equiv\prod_{i}|z_{i}\rangle$, one can
define a stochastic state vector $|\psi_{t}\rangle\equiv\langle z|\psi_{{\rm tot}}\rangle$.
Noticing the total state vector $|\psi_{{\rm tot}}\rangle$ satisfies
the Schr\"{o}dinger equation, one can obtain the dynamic equation
for $|\psi_{t}\rangle$ as 
\begin{equation}
\frac{\partial}{\partial t}|\psi_{t}\rangle=\left[-iH_{{\rm JC}}+az_{t}^{\ast}-a^{\dagger}\int\nolimits _{0}^{t}ds\alpha(t,s)\frac{\delta}{\delta z_{s}^{\ast}}\right]|\psi_{t}\rangle,\label{QSD}
\end{equation}
called the NMQSD equation \cite{Diosi1998PRA,Strunz1999PRL,Yu1999PRA}.
Equation~(\ref{QSD}) is obviously a stochastic differential equation
whose solution $|\psi_{t}\rangle$ depends on two stochastic variables.
One is the classical noise represented by stochastic function $\xi(t)$
or $\eta(t)$ in $H_{{\rm JC}}$, the other is the noise from the
quantized bath represented by the noise function $z_{t}^{\ast}=-i\sum_{i}g_{i}z_{i}^{\ast}e^{i\omega_{i}t}$.
The statistical properties of these two types of noises can be characterized
by their correlation functions
\begin{eqnarray}
M\{z_{t}\} & = & M\{z_{t}z_{s}\}=0,\nonumber \\
M\{z_{t}z_{s}^{*}\} & = & \langle\hat{B}^{\dagger}(t)B(s)\rangle=\alpha_{1}(t,s),
\end{eqnarray}
\begin{align}
\langle\xi(t)\rangle=0 & ,\quad\langle\xi(t)\xi(s)\rangle=\alpha_{2}(t,s),
\end{align}
\begin{align}
\langle\eta(t)\rangle=0 & ,\quad\langle\eta(t)\eta(s)\rangle=\alpha_{3}(t,s),
\end{align}
where $M\{\cdot\}\equiv\int\frac{dz^{2}}{\pi}e^{-|z|^{2}}\{\cdot\}$
denotes the statistical mean over the noise $z_{t}^{*}$ and $\langle\cdot\rangle$
denotes the statistical mean over the classical noises ($\xi$ or
$\eta$). The bath operator $\hat{B}(t)$ is defined as $\hat{B}(t)=\sum_{i}g_{i}b_{i}e^{-i\omega_{i}t}$.
Since the degrees of freedom of the bath is huge, the time dependent
operator $\hat{B}(t)$ can be regarded as a source randomness and
its statistical properties is governed by $\langle\hat{B}^{\dagger}(t)B(s)\rangle=\alpha_{1}(t,s)$.
The properties of these two correlation functions $\alpha_{1}(t,s)$
and $\alpha_{2}(t,s)$ {[}or $\alpha_{3}(t,s)${]} may substantially
affect the decoherence process, we will discuss their impacts in Sec.~\ref{sec:3}
and Sec.~\ref{sec:eta} in details.

It is worth to note that Eq.~(\ref{QSD}) is directly derived from
the microscopic Hamiltonian without any approximation. All the effects
(particularly the non-Markovian effects, measured by e.g., non-Markovianity
\cite{Breuer2009PRL}) in the dynamics will be captured by this equation.

The key point of solving Eq.~(\ref{QSD}) is replacing the functional
derivative $\frac{\delta}{\delta z_{s}^{\ast}}$ by a time-dependent
operator $O(t,s,z^{*})$ defined by $\frac{\delta}{\delta z_{s}^{\ast}}|\psi_{t}\rangle\equiv O(t,s,z^{*})|\psi_{t}\rangle$,
so that Eq.~(\ref{QSD}) can be rewritten as 
\begin{align}
\frac{\partial}{\partial t}|\psi_{t}\rangle & =H_{{\rm eff}}|\psi_{t}\rangle,\label{eq:QSDO}\\
H_{{\rm eff}} & =-iH_{\mathrm{JC}}+az_{t}^{\ast}-a^{\dagger}\bar{O}(t,z^{*}),
\end{align}
where $\bar{O}(t,z^{*})=\int_{0}^{t}\alpha(t,s)O(t,s,z^{*})ds$. The
operator $O$ can be determined from the consistency condition $\frac{\delta}{\delta z_{s}^{\ast}}\frac{\partial}{\partial t}|\psi_{t}\rangle=\frac{\partial}{\partial t}\frac{\delta}{\delta z_{s}^{\ast}}|\psi_{t}\rangle$
(see Appendix~\ref{app:MEQ}). Then, one can solve Eq.~(\ref{eq:QSDO})
with a single realization of the noises $\xi(t)$ {[}or $\eta(t)${]}
and $z_{t}^{*}$ to obtain a particular trajectory of the evolution.
However, the density matrix of the atom-cavity system must be reconstructed
by taking the two-fold ensemble average over many realizations of
the stochastic state vector $|\psi_{t}\rangle$, i.e., 
\begin{equation}
\rho=\left\langle M\left\{ |\psi_{t}\rangle\langle\psi_{t}|\right\} \right\rangle .\label{eq:constructRho}
\end{equation}
Here, the statistical mean $M\{\cdot\}$ (over $z_{t}^{*}$) and $\langle\cdot\rangle$
{[}over $\eta(t)$ or $\xi(t)]$ can be numerically obtained by averaging
over thousands of trajectories. Alternatively, one can also derive
a master equation by using the Novikov theorem \cite{Yu1999PRA}.
For example, in the case $\xi(t)=0$ is a constant, i.e., only the
classical noise $\eta(t)$ is applied, the master equation can be
derived as (for details, see Appendix~\ref{app:MEQ}),
\begin{equation}
\frac{d}{dt}\rho=-i\left[H_{{\rm 0}},\rho\right]+\left\{ \left[a,\rho\bar{O}^{\dagger}\right]+\left[\sigma_{z},\rho\bar{D}^{\dagger}\right]+{\rm h.c.}\right\} ,\label{eq:MEQsz}
\end{equation}
where $H_{0}=\frac{\omega_{0}}{2}\sigma_{z}+\Omega a^{\dagger}a+G_{0}\sin(kx_{0})(a\sigma_{+}+a^{\dagger}\sigma_{-})$,
the operator $\bar{D}$ is defined as $\bar{D}=i\int_{0}^{t}ds\alpha_{3}(t,s)\frac{\delta}{\delta\eta(s)}\equiv i\int_{0}^{t}ds\alpha_{3}(t,s)D(t,s,\eta)$
with the boundary condition $D(t,s=t,\eta)=\sigma_{z}$.

According to the consistency conditions $\frac{\delta}{\delta z_{s}^{\ast}}\frac{\partial}{\partial t}|\psi_{t}\rangle=\frac{\partial}{\partial t}\frac{\delta}{\delta z_{s}^{\ast}}|\psi_{t}\rangle$
and $\frac{\delta}{\delta\eta(s)}\frac{\partial}{\partial t}|\psi_{t}\rangle=\frac{\partial}{\partial t}\frac{\delta}{\delta\eta(s)}|\psi_{t}\rangle$,
one can obtain the operators $O$ and $D$ will satisfy the equations
\begin{equation}
\frac{d}{dt}D=\left[H_{{\rm eff}},D\right]+\frac{\delta}{\delta\eta(s)}H_{{\rm eff}},\label{eq:dD-1}
\end{equation}
\begin{equation}
\frac{d}{dt}O=\left[H_{{\rm eff}},O\right]+a^{\dagger}\frac{\delta}{\delta z_{s}^{*}}\bar{O},\label{eq:dO-1}
\end{equation}
with the boundary conditions $D(t=s,s,\eta)=-i\frac{1}{2}\sigma_{z}$
and $O(t=s,s,z^{*})=a$. From Eqs.~(\ref{eq:dD-1}) and (\ref{eq:dO-1}),
the operator $D$, representing the impact from noise $\eta(t)$,
depends on the $z_{t}^{*}$ term in $H_{{\rm eff}}$. In the same
way, the operator $O$, representing the impact from noise $z_{t}^{*}$
also depends on the $\eta(t)$ term. As a result, although the master
equation (\ref{eq:MEQsz}) is formally written as the summation of
two Lindblad super-operators, it does not mean the combined effect
of two noises is simply the summation of the impacts from two individual
noises. Physically, although the two noises are not correlated, they
can still affect each other through quantum system $H_{{\rm S}}$
if non-Markovian feedback effects exist. Therefore, the master equation
(\ref{eq:MEQsz}) provides a powerful tool to study the influence
of one noise on the other noise. However, it is not the central topic
of this paper and will be left for a future study.

Equations~(\ref{eq:dD-1}) and (\ref{eq:dO-1}) can be numerically
solved with iteration method \cite{Suess2014PRL}, or the operators
$O$ and $D$ can be analytically expend into series by the order
of noises up to on-demand accuracy \cite{Yu1999PRA,Li2014PRA,Xu2014JPA}.
Here, we use a simple example to show how the master equation is reduced
to the standard Lindblad master equation with constant rate of decoherence.
In the Markovian case, the correlation functions become $\delta$-functions
as $\alpha_{1}(t,s)=\Gamma_{1}\delta(t,s)$ and $\alpha_{3}(t,s)=\Gamma_{3}\delta(t,s)$.
Then, one can obtain $\bar{O}^{\dagger}=\frac{\Gamma_{1}}{2}a^{\dagger}$
and $\bar{D}=\frac{\Gamma_{3}}{2}\sigma_{z}$ from the boundary conditions
(boundary values of $O$ and $D$ are selected by the $\delta$-functions).
Therefore, the master equation in Eq.~(\ref{eq:MEQsz}) will be reduced
to the Markovian master equation in the standard Lindblad form as
\begin{equation}
\frac{d}{dt}\rho=-i\left[H_{{\rm 0}},\rho\right]+\left\{ \frac{\Gamma_{1}}{2}\left[a,\rho a^{\dagger}\right]+\frac{\Gamma_{3}}{2}\left[\sigma_{z},\rho\sigma_{z}\right]+{\rm h.c.}\right\} ,\label{eq:MEQ_Markov}
\end{equation}
The first and second Lindblad terms represent the amplitude and phase
damping of the atom respectively \cite{Gardiner2004Book}.

In a more general case with arbitrary correlation functions, the operators
$O$ and $D$ are in more complicated forms \cite{Zhao2017AoP,Zhao2011PRA,Yu1999PRA,Zhao2019OEa,Zhao2013PRA}.
It is worth to note that in order to obtain a compact form of the
master equation (\ref{eq:MEQsz}), we have assumed that both $\bar{O}$
and $\bar{D}$ are noise-independent operators. The general form of
the master equation is derived in Appendix~\ref{app:MEQ}. Nevertheless,
it is verified in Appendix~\ref{subApp:O} that the noise term is
indeed much smaller than other terms. So, it is reasonable to approximately
neglect the noise-dependent part in operators $\bar{O}$ and $\bar{D}$
to obtain a master equation in the from of Eq.~(\ref{eq:MEQsz}). 

Similarly, in the case $\eta(t)=0$, i.e., only the classical noise
$\xi(t)$ is applied, one can also derive a master equation as

\begin{equation}
\frac{d}{dt}\rho=-i\left[H_{{\rm 0}}^{\prime},\rho\right]+\left\{ \left[a,\rho\bar{O}^{\dagger}\right]+\left[V,\rho\bar{D}_{G}^{\dagger}\right]+{\rm h.c.}\right\} ,\label{eq:MEQG}
\end{equation}
where $H_{0}^{\prime}=\frac{\omega_{0}}{2}\sigma_{z}+\Omega a^{\dagger}a$,
$V=a\sigma_{+}+a^{\dagger}\sigma_{-}$, and $\bar{D}_{G}=i\int_{0}^{t}ds\left\langle G(t)G(s)\right\rangle D_{G}(t,s,G)=i\int_{0}^{t}ds\left\langle G(t)G(s)\right\rangle \frac{\delta}{\delta G(s)}$
with the boundary condition $D_{G}(t,s=t,G)=V$. The detailed derivation
is shown in Appendix~\ref{app:MEQ}. Equations~(\ref{eq:MEQsz})
and (\ref{eq:MEQG}) formally contain no convolution terms, but the
operators $O$, $D$, and $D_{G}$ include the time integration over
time, which represents the impacts from the history (non-Markovian
effects). These master equations are derived beyond the Markovian
approximation, thus applicable to either non-Markovian case or Markovian
case. When taking the Markov-limit, our equation is reduced to Markovian
equation as shown in Eq.~(\ref{eq:MEQ_Markov}). However, in non-Markovian
case, they are still valid.

For computational purpose, Eq.~(\ref{eq:QSDO}) requires the two-fold
statistical mean over many trajectories. However, the resource to
store the pure state $|\psi_{t}\rangle$ ($\propto N$, $N$ is the
dimension of the Hilbert space) is less than the resource to store
the density operator $\rho$ ($\propto N^{2}$). Therefore, when $N$
is large, the NMQSD equation has a computational advantage comparing
to the master equation Eq.~(\ref{eq:MEQsz}) or Eq.~(\ref{eq:MEQG}).
In this paper, we average the quantum noise $z_{t}^{*}$ by using
the Novikov theorem \cite{Yu1999PRA}, and obtain a statistical master
equation as 
\begin{equation}
\frac{d}{dt}\rho^{\prime}=-i\left[H_{{\rm JC}},\rho^{\prime}\right]+\left\{ \left[a,\rho^{\prime}\bar{O}^{\dagger}\right]+\mbox{H.c.}\right\} .\label{eq:MEQhalf}
\end{equation}
where $\rho^{\prime}=M\left\{ |\psi_{t}\rangle\langle\psi_{t}|\right\} $
is still a stochastic density operator only contains the classical
noise ($\eta$ or $\xi$ in $H_{{\rm JC}}$). In order to obtain the
density operator $\rho$, one also need to numerically take the statistical
mean over the classical noise $\eta(t)$ or $\xi(t)$ as $\rho=\langle\rho^{\prime}\rangle$.
Equation~(\ref{eq:MEQhalf}) has a unified format for the two types
of classical noises $\eta(t)$ or $\xi(t)$. Meanwhile, it only requires
taking a single-fold average. Therefore, it is a balanced choice between
Eq.~(\ref{eq:QSDO}) and Eq.~(\ref{eq:MEQsz}) {[}or Eq.~(\ref{eq:MEQG}){]}.
The former one requires two-fold average and the latter ones do not
in a unified format for different noises $\eta(t)$ or $\xi(t)$.

\section{Coherence protection by noisy}

\label{sec:3}

The fundamental dynamical equations are derived as Eqs.~(\ref{eq:QSDO},
\ref{eq:MEQsz}, \ref{eq:MEQG}, \ref{eq:MEQhalf}) in the last section.
Based on these equations, we will focus on the protection of coherence
in this section. From the derivation of Eq.~(\ref{QSD}), $\alpha_{1}(t,s)=\sum_{i}|g_{i}|^{2}e^{-i\omega_{i}(t-s)}=\int_{0}^{\infty}g(\omega)e^{-i\omega(t-s)}d\omega$
can be interpreted as a Fourier transformation of the spectrum density
$g(\omega)$. In the numerical simulation, we choose the Ornstein-Uhlenbeck
(O-U) correlation function for all the three noises
\begin{equation}
\alpha_{i}(t,s)=\frac{\Gamma_{i}\gamma_{i}}{2}e^{-\gamma_{i}|t-s|},\quad(i=1,2,3),\label{eq:OUN1}
\end{equation}
corresponding to a spectrum density in the Lorentzian form \cite{Breuer2009PRL,Tu2008PRB}
\begin{equation}
g_{i}(\omega)=\frac{1}{2\pi}\frac{\Gamma_{i}\gamma_{i}^{2}}{\omega^{2}+\gamma_{i}^{2}},\quad(i=1,2,3).\label{eq:gw}
\end{equation}
The reason to choose such a correlation function is to clearly observe
the transition from Markovian regime to non-Markovian regime, since
the memory time of the noise is explicitly indicated by the parameter
$1/\gamma_{i}$. When $\gamma_{i}\rightarrow\infty$, $\alpha_{i}(t,s)\rightarrow\Gamma_{i}\delta(t,s)$,
it is reduced to the Markovian correlation function. Besides, other
types of correlation functions can be also decomposed into combinations
of several O-U correlation functions due to the fact that an arbitrary
function can be expanded into exponential Fourier series. Examples
of using arbitrary correlation functions are shown in Appendix~\ref{app:OUnoise},
where the widely used $1/f$ noise is decomposed into many O-U noises
and an example of telegraph noise is also given.

In this section, we will investigate how the properties of both the
classical noise $\xi(t)$ {[}or $\eta(t)${]} and the quantum noise
$z_{t}^{*}$ affect the performance of the protection. Particularly,
we find several interesting phenomena caused by the memory effects
of the two noises. The numerical studies show the performance of the
coherence protection sensitively depends on the properties of both
noises, particularly the memory effects of the two noises.

\subsection{\label{subsec:Mechanism}Mechanism of coherence protection}

In order to understand the mechanism of the coherence protection,
we analyze the dynamics of the atomic coherence in a simplified case
that the total excitation in the atom and the cavity is limited to
1. In this case, a general total state vector can be written in a
subspace as

\begin{eqnarray}
|\psi_{{\rm tot}}(t)\rangle & = & A(t)|e,0,0_{{\rm B}}\rangle+B(t)|g,1,0_{{\rm B}}\rangle+C(t)|g,0,0_{{\rm B}}\rangle\nonumber \\
 & + & \sum_{k}D_{k}(t)|g,0,1_{k}\rangle,\label{eq:GenVec}
\end{eqnarray}
where $|0\rangle$ and $|1\rangle$ are the Fock states of the cavity,
$|0_{B}\rangle$ is the collective vacuum state of the bath, and $|1_{k}\rangle$
is the first excited state of $k^{{\rm th}}$ mode. Substituting the
state $|\psi_{{\rm tot}}(t)\rangle$ into the Schrödinger equation
$-i\frac{d}{dt}|\psi_{{\rm tot}}(t)\rangle=H_{{\rm tot}}|\psi_{{\rm tot}}(t)\rangle$,
one can obtain the a set of dynamical equations 
\begin{eqnarray}
\frac{d}{dt}A(t) & = & -iG(t)B(t),\label{eq:dA}\\
\frac{d}{dt}B(t) & = & -iG(t)A(t)-I(t),\label{eq:dB}\\
\frac{d}{dt}I(t) & = & \frac{\Gamma_{1}\gamma_{1}}{2}B(t)-\gamma_{1}I(t),\label{eq:dI}
\end{eqnarray}
where $I(t)=\int_{0}^{t}\frac{\Gamma_{1}\gamma_{1}}{d}e^{-\gamma_{1}(t-s)}B(s)ds$,
$A(0)=\frac{1}{\sqrt{2}}$ is finite, and $B(0)=0$, and the coefficient
$C(t)=C(0)$ will not change during the evolution. The atomic coherence
(off-diagonal elements of the reduced density matrix) can be expressed
as
\begin{equation}
|\rho_{a}(1,2)|=|AC^{*}|.\label{eq:rhoa12}
\end{equation}
Integrating Eq.~(\ref{eq:dA}), one can obtain
\begin{equation}
A(t)=A(0)-i\int_{0}^{t}G(s)B(s)ds.\label{eq:Aint}
\end{equation}
Assuming $G(t)$ is varying much faster than $B(t)$, the time-dependent
function $B(t)$ can be treated as time-independent in the integration.
Then, $A(t)$ is determined by the integration $\int_{0}^{t}G(s)ds=\int_{0}^{t}G_{0}\sin\left\{ k\left[x_{0}+\xi(t)\right]\right\} ds.$
The average effect will be zero when the standard deviation of $k\xi(t)\sim\pi$,
and $A(t)$ is frozen to $A(0)$. So, the quantum state will be ``freeze''
to the initial with quantum coherence being protected.

The varying speed of $B(t)$ is actually determined by $I(t)$ from
Eq.~(\ref{eq:dB}), because the contribution of $A(t)$ is also zero
in the integration for the same reason if $G(t)$ is a fast varying
function. According to Eq.~(\ref{eq:dA}-\ref{eq:dI}), $I(t)$ is
directly determined by the properties of the noise $z_{t}^{*}$, because
$\gamma_{1}$ and $\Gamma_{1}$ appear in the differential equation
(\ref{eq:dI}). Therefore, the mathematical condition $G(t)$ is varying
much faster than $B(t)$ can be somehow interpreted as a physical
condition that the noise $\xi(t)$ is much choppier than the noise
$z_{t}^{*}$. Although the discussion above is based on a simplified
example, it capture the main picture of the coherence protection.
It is worth to note the that a general case should be subjected to
Eqs.~(\ref{eq:QSDO}, \ref{eq:MEQsz}, \ref{eq:MEQG}, \ref{eq:MEQhalf}),
where the mechanism should be different but similar (see Ref.~\cite{jing2013SR}
as an example).

In the analysis above, we have shown the mechanism of the coherence
protection from the mathematical perspective. In a brief summary,
in order to eliminate the negative impact from a low-frequency noise,
one can appropriately introduce another high-frequency noise, eliminating
the contribution of the low-frequency part in the time-integral. Interestingly,
the trouble caused by a noise is suppressed by another choppier noise.
Actually, the well known coherence protection schemes such as dynamical
decoupling \cite{Vedral1997PRL,Viola1998PRA,Viola1999PRL} or Zeno
effect \cite{Facchi2004PRA,Misra1977JoMP,Ai2010PRA} are based on
the similar mathematical reasons. A stochastic coupling may randomly
reverse the coupling, this may isolate the quantum system to its environment.
Here, we replace the artificially designed operations (pulses or measurements)
by noise and obtain the similar effect of coherence protection.

\subsection{Non-Markovian noise $\xi(t)$ in the coupling}

\begin{figure*}[t]
\begin{centering}
\includegraphics[width=0.9\textwidth]{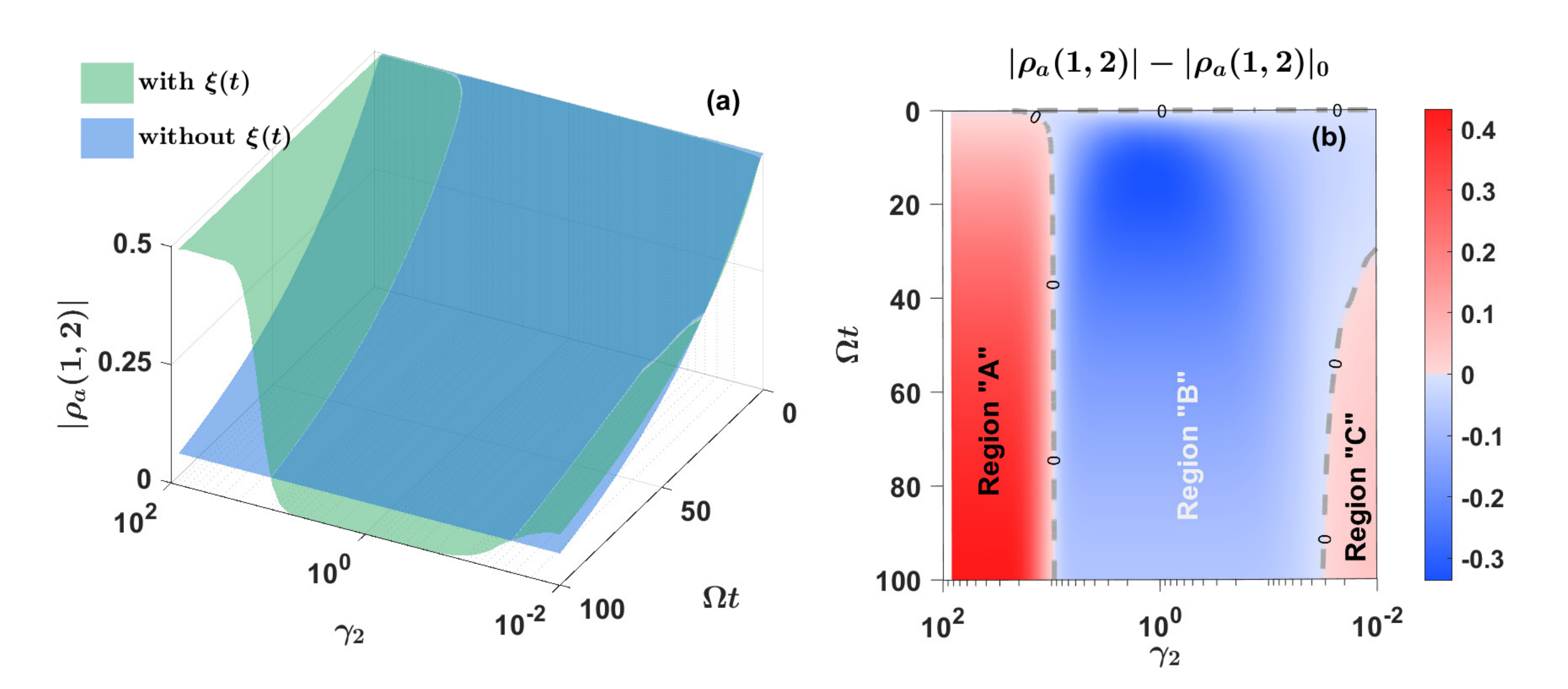}
\par\end{centering}
\caption{\label{fig:Cr2t}(a) Time evolution of the atomic coherence ($|\rho_{a}(1,2)|$,
$\rho_{a}=\mbox{Tr}_{c}\{\rho_{ac}\}$) under different memory time
$1/\gamma_{2}$ with noise and without noise of $\xi(t)$. The parameters
are $\omega=\Omega=1$, $kx_{0}=0.08$, $\gamma_{1}=1$, and $\Gamma_{1}=\Gamma_{2}=1$.
(b) Difference of the coherence between the cases with and without
$\xi(t)$. $|\rho_{a}(1,2)|$ indicates the coherence with noise and
$|\rho_{a}(1,2)|_{0}$ indicates the coherence without noise.}
\end{figure*}
In this subsection, we will use numerical results to illustrate the
mechanism proposed in Sec.~\ref{subsec:Mechanism} and show how to
use non-Markovian behaviors to protect coherence. We start from the
noise $\xi(t)$ with a finite memory time $1/\gamma_{2}$. The time
evolution of the atomic coherence with and without the noise $\xi(t)$
is plotted in Fig.~\ref{fig:Cr2t}~(a). The green surface indicates
the coherence evolution in presence of the noise $\xi(t)$ with various
$\gamma_{2}$, while the blue surface indicating the coherence evolution
without noise. In order to show the effect of protection clearly,
we also plot the difference of these two surfaces in Fig.~\ref{fig:Cr2t}~(b),
where the red color indicates the noise $\xi(t)$ has a positive effect
on coherence protection and the blue color indicates a negative effect.
One can also check the purity $P={\rm Tr}(\rho_{a}^{2})$ (characterize
the state is pure or not) follows the similar pattern at the early
stage of the evolution (not shown). It is shown in Fig.~\ref{fig:Cr2t}
that the coherence protection highly depends on the memory time $1/\gamma_{2}$
of the noise $\xi(t)$. When $\gamma_{2}$ is above a threshold (region
``A''), one can observe very strong protection of the atomic coherence.
In contrast, below this threshold, the coherence loss is even faster
(region ``B''). However, at the right-bottom corner, there is a
smaller region ``C'' with a small but positive effect of coherence
protection.

Recall the mechanism proposed in Sec.~\ref{subsec:Mechanism}, since
the coherence protection requires a fast varying speed of $G(t)$,
the noise $\xi(t)$ should contain high-frequency components. The
spectrum density in Eq.~(\ref{eq:gw}) shows that $g_{2}(\omega)$
has a finite high-frequency distribution only if $\gamma_{2}$ is
large enough. Otherwise, when $\gamma_{2}\rightarrow0$, $g_{2}(\omega)\propto\omega^{-2}$,
the high-frequency components will quickly decrease to zero when $\omega$
is increasing. Physically, a Markovian noise $\xi(t)$ with $\gamma_{2}\rightarrow\infty$
has a uniform distribution on all the frequencies (white noise), corresponding
to a $\delta(t,s)$ correlation function. Those high-frequency noises
will be benefit to the coherence protection. Therefore, in Fig.~\ref{fig:Cr2t},
the Markovian noise ($\gamma_{2}\rightarrow\infty$) with more high-frequency
components is more useful to the coherence protection, reflected by
positive effect in region ``A''. In contrast, in non-Markovian case,
the frequency distribution of $\xi(t)$ is mainly concentrated in
the low-frequency regime {[}see Eq.~(\ref{eq:gw}){]}. The coupling
$G(t)$ in Eq.~(\ref{eq:dA}) is not varying faster than $B(t)$,
then $B(t)$ can not be treated as constant in the integral and its
impact can not be eliminated. Even worse it may introduce extra loss
of coherence as shown in the region ``B'' in Fig.~\ref{fig:Cr2t}~(b).
However, we also notice there is another positive region ``C'' when
$\gamma_{2}\rightarrow0$, this is a different mechanism of averaging
$\xi(t)$.

\noindent 
\begin{figure}
\begin{centering}
\includegraphics[width=1\columnwidth]{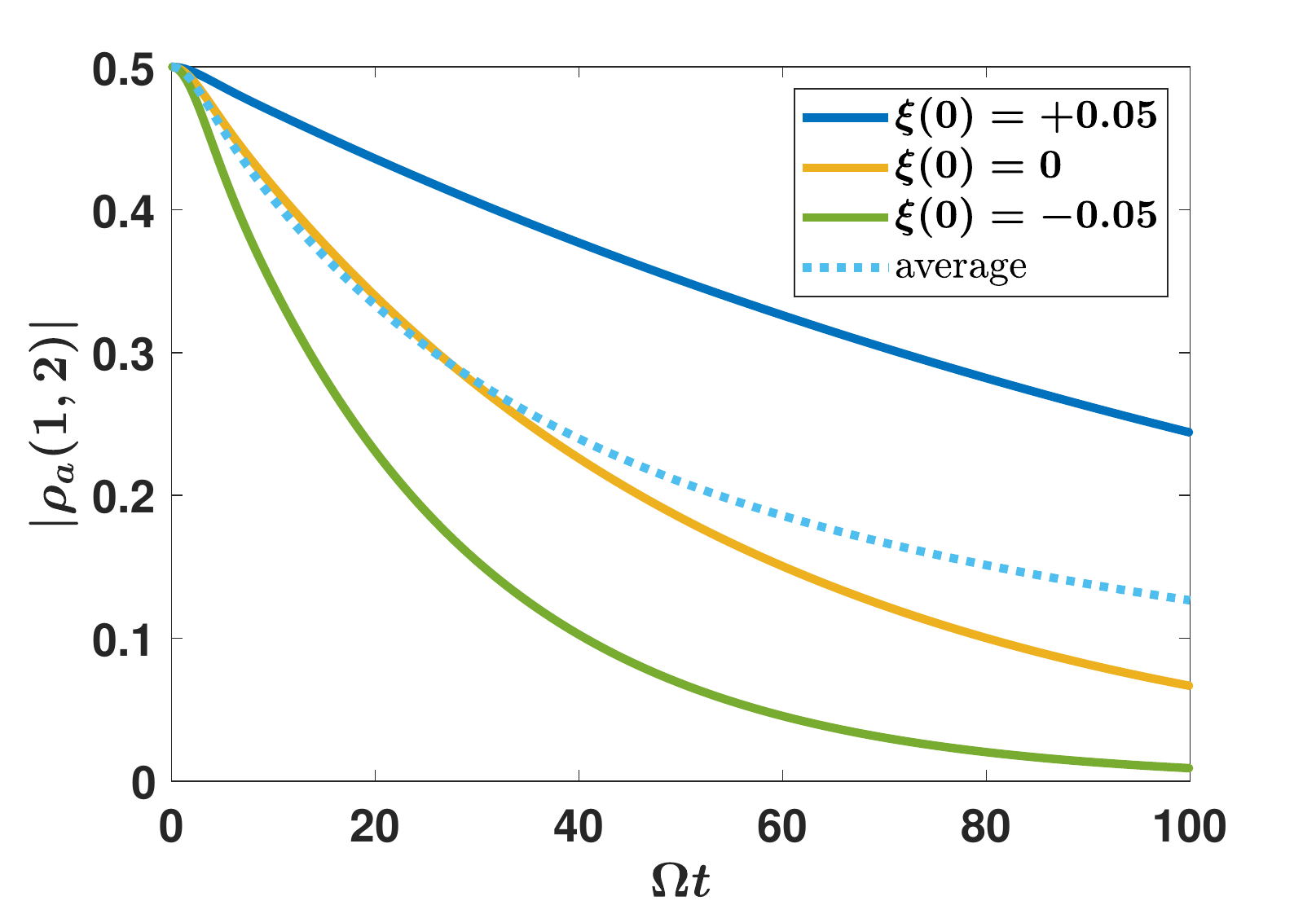}
\par\end{centering}
\caption{\label{fig:xi0}Coherence protection by super-long memory time of
$\xi(t)$. The parameter is the same as Fig.~\ref{fig:Cr2t}.}
\end{figure}

When $\gamma_{2}\rightarrow0$, the memory time of the noise $\xi(t)$
is even longer than the evolution time. During the evolution, $\xi(t)$
can be approximately treated as ``not varying'', namely $\xi(t)=\xi(0)$.
Therefore, for a particular trajectory, the time evolution is only
determined by the initial value $\xi(0)$. For a particular realization,
if $\xi(0)$ is randomly chosen as a positive number $\xi(0)=+c$,
it will accelerate the dissipation. On the contrary, if $\xi(0)=-c$,
it will decrease the dissipation. The average over the noise $\xi(t)$
is actually an average of these two effects. An example is shown in
Fig.~\ref{fig:xi0}, where the dotted curve labeled with ``average''
is the average of the two curves $\xi(0)=+0.05$ and $\xi(0)=-0.05$.
The curve $\xi(0)=0$ can be regarded as the case without noise $\xi(t)$.
When adding the noise $\xi(t)$, the average effect lead to a higher
residue coherence at $\Omega t=100$. At the early stage of the evolution
(approximately $\Omega t<30$), the average causes a negative effect.
This is in accordance with the results shown in Fig.~\ref{fig:Cr2t},
where the coherence protection has a negative effect at the early
stage of the evolution at $\gamma_{2}\rightarrow0$.

\subsection{Combined effects of two noises}

\noindent 
\begin{figure*}
\begin{centering}
\includegraphics[width=1\linewidth]{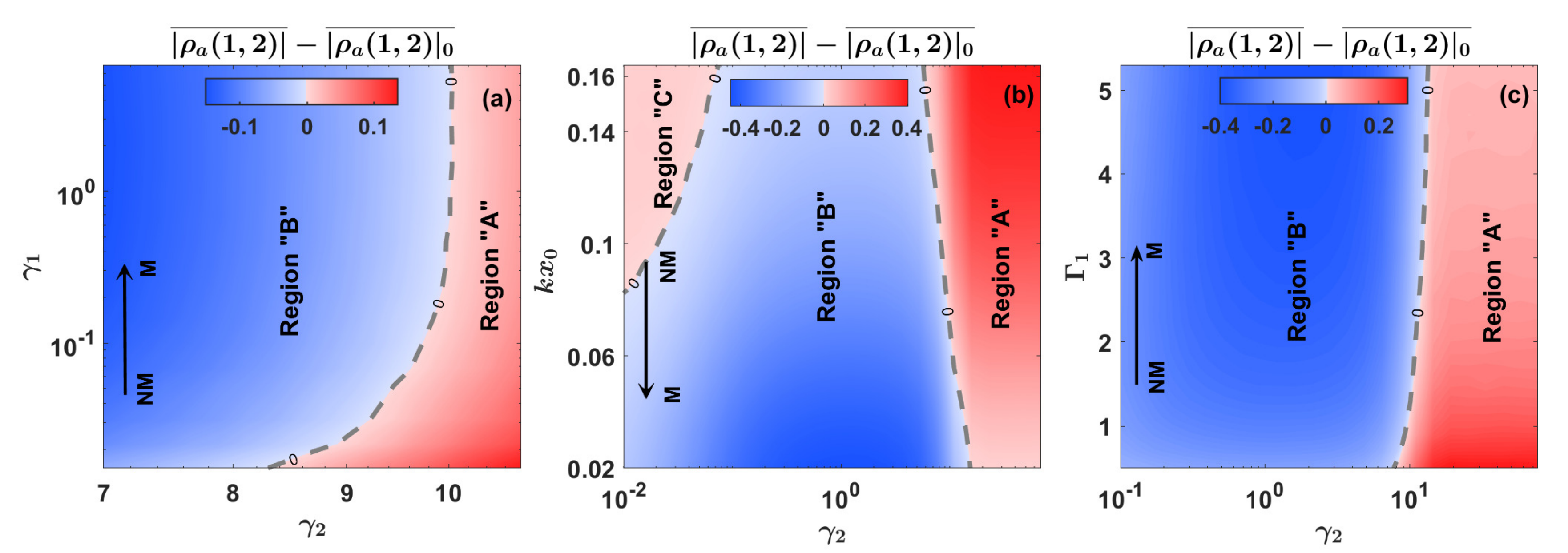} 
\par\end{centering}
\caption{\label{fig4:r2vs3factor}Coherence protection: impacts from two noises.
The performance is characterized by the difference of average coherence
of $\overline{|\rho_{a}(1,2)|}=\frac{1}{\Omega T}\int_{0}^{\Omega T=100}|\rho_{a}(1,2)|(t)dt$
{[}with $\xi(t)${]} and $\overline{|\rho_{a}(1,2)|_{0}}$ {[}without
$\xi(t)${]}. With the change of the factors indicated by the $y$-axis
of each sub-plot, the arrows show the direction of transition from
non-Markovian (NM) to Markovian (M) regime for the quantized noise
$z_{t}^{*}$. The parameters are $\omega_{0}=\Omega=1$, $\gamma_{1}=0.5$,
$kx_{0}=0.08$, $G_{0}=1$, and $\Gamma_{1}=\Gamma_{2}=1$, unless
specified explicitly in each sub-plot.}
\end{figure*}

In the last subsection, we focus on the non-Markovian effect of $\xi(t)$
and analyze the mechanism of the coherence protection. Besides the
noise $\xi(t)$, there is also another quantized noise $z_{t}^{*}$
from the bosonic bath $H_{\mathrm{B}}=\sum_{i}\omega_{i}b^{\dagger}b_{i}$.
From the view of the atom, the cavity plus the bath is a hierarchical
environment, whose effective correlation function is complicated \cite{Mazzola2009PRA}.
Here, we only make a simple qualitative analysis. The non-Markovianity
of the hierarchical environment is determined by several factors.

First, a longer memory time $1/\gamma_{1}$ certainly leads to a non-Markovian
effect, although such an impact is indirectly transmitted to the atom
through the cavity. In Fig.~\ref{fig4:r2vs3factor}~(a), when $\gamma_{1}$
is small, the hierarchical environment is in the non-Markovian regime.
Typically, high-frequency components is weak in non-Markovian noise,
and Markovian noise contains more high-frequency components. Taking
the O-U noise in Eq.~(\ref{eq:OUN1}) as an example, when $\gamma_{1}$
is small, the high-frequency part in the spectrum density is suppressed
as shown in Eq.~(\ref{eq:gw}). As we have analyzed, the mechanism
of coherence protection is the low-frequency noise can be controlled
by another high-frequency noise. If a non-Markovian noise $z_{t}^{*}$
(long memory time) causes a slow evolution of the system, we only
need to use a moderately high-frequency $\xi(t)$ to cancel the impact
of $z_{t}^{*}$. It is reflected in Fig.~\ref{fig4:r2vs3factor}~(a)
as the threshold of $\gamma_{2}$ to achieve positive effect of protection
is relatively low when $\gamma_{1}$ is small. On the contrary, when
$\gamma_{1}$ is large, $z_{t}^{*}$ is a Morkov noise, containing
more high-frequency components. Then, we need to use a much higher
frequency noise to suppress it. Reflected in Fig.~\ref{fig4:r2vs3factor}~(a),
the threshold becomes larger when $\gamma_{1}$ is increased. 

Second, the coupling strength $G$ is also crucial to the non-Markovianity
of the noise $z_{t}^{*}$. One one hand, taking the cavity plus the
bath as combined environment, a stronger coupling to the system certainly
causes stronger non-Markovian feedback effect. On the other hand,
the cavity (pseudo-environment) can be regarded as a tiny reservoir,
the ``water level'' in the reservoir is determined by both the injection
rate $G$ and the leakage rate $\Gamma_{1}$. For a fixed $\Gamma_{1}$,
a smaller $G$ will cause all the ``water'' is drained. Certainly,
there will be no feedback effect (backflow). In our case, although
the coupling $G(t)$ is a stochastic function, the balanced position
$x_{0}$ still indicates an average effect. Without the second noise
$\xi(t)$, a larger $x_{0}$ ($kx_{0}<\pi/2$) means a stronger coupling,
namely a stronger non-Markovian effect of $z_{t}^{*}$. Then, as we
have analyzed above, a moderately Markovian (high-frequency) noise
$\xi(t)$ can successfully suppress $z_{t}^{*}$. On the contrary,
a small $x_{0}$ corresponds to a Markovian evolution. Thus, one need
a noise $\xi(t)$ with a higher frequency (more Markovian) with a
larger $\gamma_{2}$. This is illustrated by the numerical result
in Fig.~\ref{fig4:r2vs3factor}~(b). When $kx_{0}$ is decreased
(noise $z_{t}^{*}$ is changing from non-Markovian to Markovian),
the threshold to achieve positive effect of coherence protection is
increased. As for the anomalous region ``C'', the reason for a positive
effect is the same as we have discussed in Fig.~\ref{fig:xi0}.

Third, the dissipation rate to the bath $\Gamma_{1}$ can also determine
the non-Markovianity of $z_{t}^{*}$. When the cavity is treated as
the ``reservoir'', the a small dissipation rate leads to a strong
backflow to the system. In the limiting case, if $\Gamma_{1}\rightarrow0$,
the total environment becomes a single cavity with a single mode.
This is definitely a very strong non-Markovian environment. Therefore,
it is shown in Fig.~\ref{fig4:r2vs3factor}~(c) that with the increase
of $\Gamma_{1}$, the threshold of achieving positive effect is also
increased.

In a brief summary, we use the numerical results to illustrate that
a low-frequency noise can be suppressed by a high-frequency noise.
This provides a clue to engineer the environment \cite{Liu2011NP}
so that making the combined decoherence effect of two noises is reduced.
The question (ii) raised in Sec.~\ref{sec:Intorduction} is partially
answered from the perspective of non-Markovian properties of the noises.

\subsection{\label{sec:eta}Non-Markovian noise $\eta(t)$}

As we have mentioned in Sec.~\ref{sec:Intorduction}, in the topic
of coherence protection by noise, one important question is how to
add the second noise to make the effect of protection is positive.
Besides the noise $\xi(t)$ in coupling, another noise is often appear
as the dephasing noise for the atom. Particularly, in the cases discussed
Appendix~\ref{app:DQD}, the noise the external field often caused
a fluctuation in the energy gap of the qubit, represented by $\eta(t)$
in Eq.~(\ref{eq:wt}). In this subsection, we will mainly focus on
this type of noise.

In order to compare with the case of $\xi(t)$, we rotate the Hamiltonian
with respect to $S=\frac{\Phi(t)}{2}\sigma_{z}$, where $\Phi(t)=\int_{0}^{t}\eta(s)ds$.
The Hamiltonian in this rotating frame becomes
\begin{align}
H_{{\rm JC}}^{\prime} & =e^{iS}H_{{\rm JC}}e^{-iS}-\frac{\partial S}{\partial t}\nonumber \\
 & =\frac{\omega_{0}}{2}\sigma_{z}+\Omega a^{\dagger}a+G_{0}e^{i\Phi(t)}a\sigma_{+}+G_{0}e^{-i\Phi(t)}a^{\dagger}\sigma_{-}.\label{eq:HJCp}
\end{align}
The original noise in $\omega(t)=\omega_{0}+\eta(t)$ has been transformed
into a phase noise $\Phi(t)$ in the coupling. Moreover, different
from the stochastic coupling $G(t)=G_{0}\sin\left\{ k\left[x_{0}+\xi(t)\right]\right\} $,
the stochastic coupling $G_{0}e^{i\Phi(t)}$ has a fixed amplitude,
but a random phase factor. This will cause slightly difference on
the coherence protection as shown below.

\noindent 
\begin{figure*}
\begin{centering}
\includegraphics[width=0.9\textwidth]{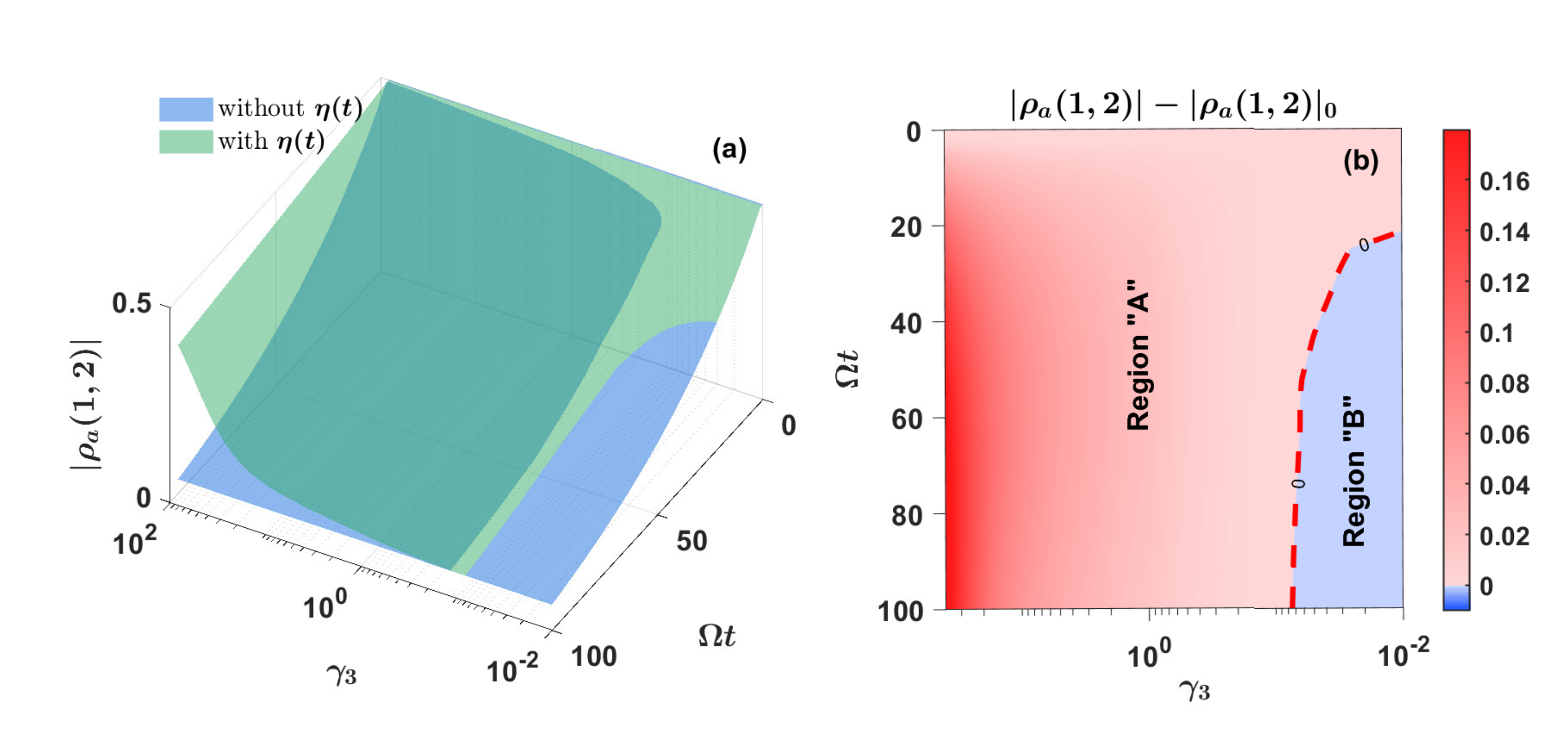}
\par\end{centering}
\caption{\label{fig:szr2t}Time evolution of the atomic coherence ($|\rho_{a}(1,2)|$)
under different memory time $1/\gamma_{3}$ with and without noise
of $\eta(t)$. The parameters are $\omega_{0}=\Omega=1$, $G_{0}=0.1$,
$\gamma_{1}=1$, and $\Gamma_{1}=\Gamma_{2}=1$. (b) Difference of
the coherence between the cases with and without $\eta(t)$. $|\rho_{a}(1,2)|_{0}$
indicates the coherence without noise.}
\end{figure*}

In Fig.~\ref{fig:szr2t}, we investigate the impact of the memory
time $\eta(t)$. Similar to the case of $\xi(t)$, the coherence protection
has a positive effect in the Markovian regime. According to the mechanism
of coherence protection discussed in Sec.~\ref{subsec:Mechanism},
a high-frequency noise is more welcome, so that region ``A'' is
positive. However, different from the special region ``C'' in Fig.~\ref{fig:Cr2t}~(b),
the positive region at $\gamma_{3}\rightarrow0$ does not appear in
Fig.~\ref{fig:szr2t}. This is because the average mechanism shown
in Fig.~\ref{fig:xi0} is not applicable to the Hamiltonian (\ref{eq:HJCp}).
The noise $\xi(t)$ causes a fluctuation around the balanced value
$G_{0}\sin(kx_{0})$, so the amplitude of $G$ is fluctuating. As
comparison, the noise $\eta(t)$ only changes the phase of the coupling
$G$ without changing its amplitude. Mathematically, $\gamma_{3}\rightarrow0$
leads to $G(t)\rightarrow G_{0}e^{\pm i\phi_{c}}$ ($\phi_{c}$ is
a constant phase factor due to super-long memory time) in Eq.~(\ref{eq:dA}-\ref{eq:dI}).
Then, it is straightforward to check that adding a constant phase
factor only results in a phase factor $e^{\pm i\phi_{c}}$ on the
solution $A(t)$ as $A(t)\rightarrow A(t)e^{\pm i\phi_{c}}$. This
has no contribution to the coherence in Eq.~(\ref{eq:rhoa12}). Therefore,
in the limit of $\gamma_{3}\rightarrow0$, we see $|\rho_{a}(1,2)|$
is asymptotically approaching $|\rho_{a}(1,2)|_{0}$. The positive
region like region ``C'' in Fig.~\ref{fig:Cr2t}~(b) at $\gamma_{3}\rightarrow0$
does not appear in Fig.~\ref{fig:szr2t}.

From this example, we see the form of the second noise is also crucial
to the coherence protection. We would like to emphasize that all the
results shown in this section only imply that the quantized noise
$z_{t}^{*}$ can be suppressed by $\text{\ensuremath{\eta(t)}}$ or
$\xi(t)$. It is also possible to show the opposite case that by comparing
the coherence with and without $z_{t}^{*}$, the impacts of a classical
noise can be weaken, too. Certainly, the condition to achieve that
goal must be different (maybe the coupling form is different from
$\eta(t),$or the required spectrum density is different). Certainly,
all of these topics are valuable in another study elsewhere.

\subsection{Outlook: Alternative perspective of coherence protection}

\noindent 
\begin{figure}
\begin{centering}
\includegraphics[width=1\columnwidth]{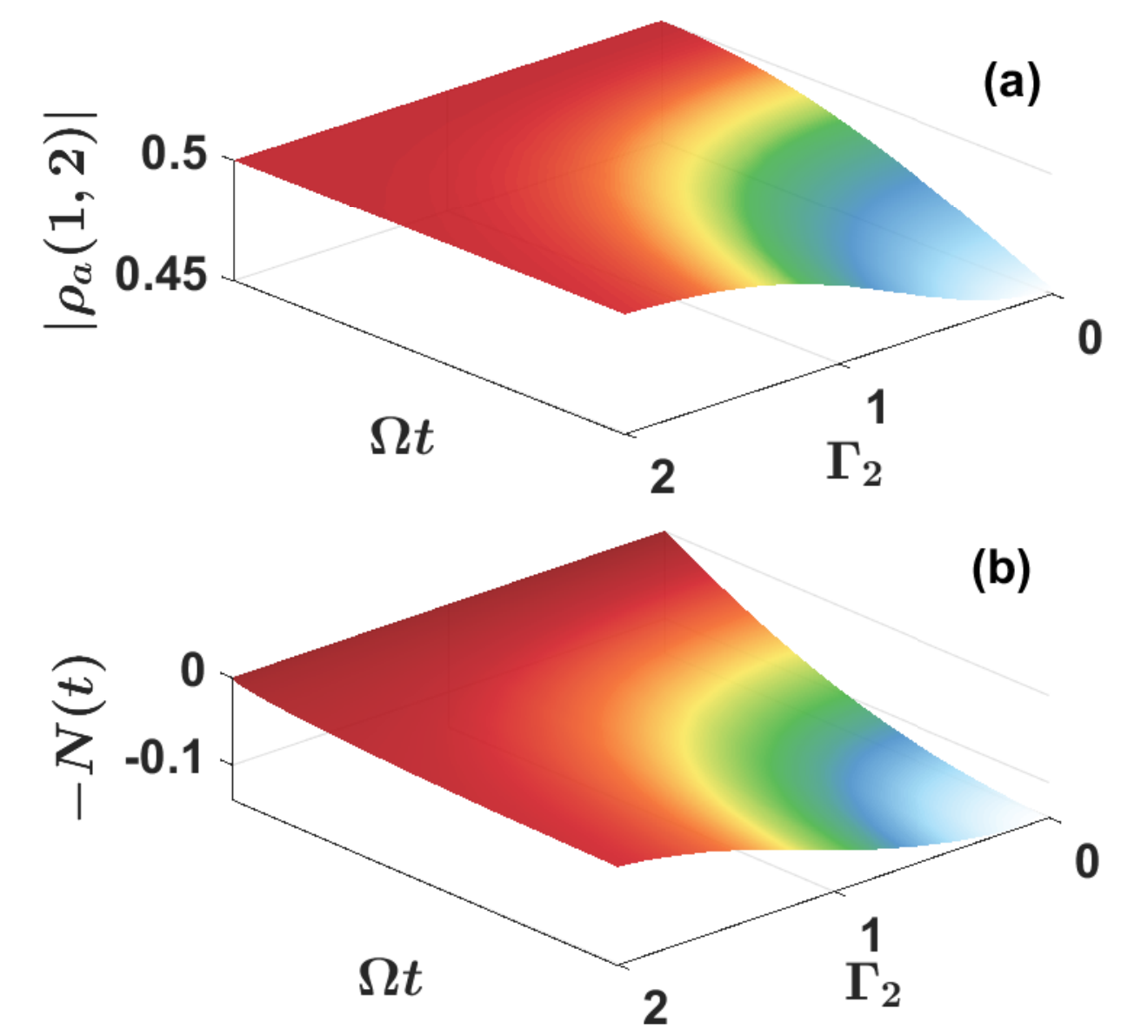}
\par\end{centering}
\caption{\label{fig:CohNeg}Time evolution of the atomic coherence $|\rho_{a}(1,2)|$
vs. the atom-cavity entanglement $N(t)$ \cite{Vidal2002PRA,Vidal2000JoMO}.
To compare with the coherence, a minus sign is added artificially
as {[}$-N(t)${]}, lower values represent stronger entanglement. The
parameters for all the sub-plots are $\omega=\Omega=1$, $\Gamma_{1}=0.5$,
$\gamma_{1}=1$, and $x_{0}=0.1$.}
\end{figure}

In the previous subsections, we have made an extensive investigation
on coherence protection by noise. At last, we would like to analyze
the coherence protection in a different perspective and it is our
hope that it may shed new light on answering the question (i) raised
in Sec.~\ref{sec:Intorduction}.

Besides the mathematical picture, we would also analyze the coherence
protection from a physical perspective. Physically, there are mainly
three reasons causing the loss of coherence in our model, (1) atom-cavity
entanglement, (2) atom-bath entanglement, and (3) energy dissipation
to the bath. These three processes happen at the same time in the
evolution and the coherence loss is a combined effect. First, the
decay rate to the bath $H_{{\rm B}}$ is proportional to the photon
number $|B(t)|^{2}$ as 
\begin{equation}
\frac{d}{dt}|D_{k}(t)|^{2}=-|g_{k}|^{2}|B(t)|^{2}.
\end{equation}
In order to reduce the decay rate to the bath, one need to suppress
the photon number $|B(t)|^{2}$ in the cavity. However, in this particular
model, reducing the photon number $|B|^{2}$ happens to also reduce
the entanglement between the atom and the cavity, since the atom-cavity
entanglement can be measured by the concurrence \cite{Wootters1998PRL}
as 
\begin{equation}
C(\rho_{ac})=4|A|^{2}|B|^{2}.\label{eq:Crho}
\end{equation}
In Eq.~(\ref{eq:dB}), the changing of $B(t)$ at the early stage
of the evolution mainly comes from the contribution of $-iG(t)A(t)$,
because $I(t)$ is a slow varying term. We have already shown that
the average of a noisy $G(t)$ is close to zero in the time integral.
Therefore, a slow increasing of the photon number $|B(t)|^{2}$ at
the early stage of the evolution can prevent not only the decay to
the bath but also the entanglement generation.

In this particular model, the atom-cavity entanglement generation
is directly related the coherence protection {[}see Eq.~(\ref{eq:rhoa12})
and Eq.~(\ref{eq:Crho}){]}. In Fig.~\ref{fig:CohNeg}, we numerically
show the connection between coherence protection and the atom-cavity
entanglement generation. Roughly, the patterns of the dynamics are
almost identical. For more details of the numerical simulation, see
Appendix~\ref{app:outlook}. 

In a more general picture shown in Fig.~\ref{fig:1}, the cavity
here can be regarded as part of hierarchical environment. It is known
that a quantum system will lose its coherence when taking the trace
of the environmental degrees of freedom, if the quantum system is
entangled with the environment. For this particular example, we see
that using a noise to destroy the entanglement generation can protect
the coherence. In a more general case with a hierarchical environment,
this might be a clue to find what types of the second noise has a
positive effect on coherence protection and partially answer the question
(i) raised in Sec.~\ref{sec:Intorduction}.

Interestingly, From the perspective of destroying the atom-cavity
entanglement, the anomalous pattern in Fig.~\ref{fig:Cr2t} can be
also explained by non-Markovian effect on entanglement generation.
Markovian noise corresponding to a larger $\gamma_{2}$ is more powerful
to destroy the atom-cavity entanglement generation thus the coherence
is protected (region ``A''). In contrast, non-Markovian noise corresponding
to a smaller $\gamma_{2}$ is often considered to be good to the entanglement
generation \cite{Ding2021SR,Zhao2019OEa,Zhao2011PRA,Zhao2012PRA,Shen2017PRA,Shen2018PRA,Shen2018OL,Shen2018PRAb},
so that the coherence is destroyed (region ``B''). The interesting
phenomenon arises in region ``C'' in Fig.~\ref{fig:Cr2t}~(b)
can be also explained by the entanglement generation in non-Markovian
dynamics, which has been widely studied in several references \cite{Zhao2011PRA,Zhao2012PRA,Zhao2019OEa,Zhao2017AoP}.
For example, recall the results in Refs. \cite{Zhao2011PRA,Zhao2019OEa},
the entanglement generated in a non-Markovian environment characterized
by O-U correlation function reaches the maximum value when the memory
time is neither too big nor too small. Therefore, the minimum coherence
should appear when $\gamma_{2}$ is neither too big nor too small.
Reflected in Fig.~\ref{fig:Cr2t}~(b), there is a ``negative effect''
window opened between two ``positive effect'' regions.

Certainly, all the analysis above is only based on a very special
case, and the connection between entanglement generation and coherence
protection is far more complicated than we discussed here. But, we
would like to emphasize that a future research on this topic may open
a new path to design more coherence protection schemes by using noise.

\section{Discussion and Conclusion}

\label{sec:5} In this paper, we show how to protect the quantum coherence
by using noise. As an example, we solve the J-C model coupled to an
external environment and derive the master equation incorporating
the non-Markovian behaviors from both the classical noise and the
quantized bath. The master equation itself is valuable in theoretical
study on the joint non-Markovian effect of two noises. It may also
become a powerful tool to investigate the indirectly interaction of
the two noises in future researches. Beyond the theoretical significance,
we also show that the atomic coherence can be protected by adding
another noise. By analyzing the non-Markovian properties of the two
noises, we draw a conclusion that the decoherence caused by a low-frequency
noise can be eliminated by another noise containing higher-frequency
components. The results show that the non-Markovian properties (particularly
the memory time) of the two noises are crucial to the performance
of the coherence protection. At last, as an outlook, we also discuss
the relation between the atom-cavity entanglement and the atomic coherence.

This work may provide a deeper understanding of noise. In most cases,
noise is often supposed to be harmful to quantum coherence. However,
the example discussed in this paper illustrates that noise can be
helpful to maintain quantum coherence in certain cases. Similar to
several other studies \cite{Yi2003PRA,jing2013SR,Jing2014PRA,Jing2018PRA,Khodorkovsky2008PRL,Chen2017PRA},
the results presented in this paper provide an example once again
that noise can lead not only troubles but also benefits. This may
shed more light on the understanding of noise. Besides, it is very
interesting that the trouble caused by one noise ($z_{t}^{*}$) happens
to be eliminated by another noise ($\eta$ or $\xi$). Similar results
have been observed in several other theoretical studies of different
models \cite{jing2013SR,Jing2018PRA,Qiao2019SCPM&A}. Here, we analyze
the reason for the coherence protection from two aspects. Mathematically,
a slow-varying function can be eliminating in the time integral by
a fast varying function. Physically, the relation between entangling
to the environment and the loss of coherence inspire us to search
a possible way to protect coherence by preventing it to be entangled
with the environment. This may open a new path to design new coherence
protection schemes by using noise.
\begin{acknowledgments}
This work was supported by the National Natural Science Foundation
of China under Grants No. 11575045, the Natural Science Funds for
Distinguished Young Scholar of Fujian Province under Grant No. 2020J06011,
Project from Fuzhou University under Grant No. JG202001-2, Project
from Fuzhou University under Grant No. GXRC-21014.
\end{acknowledgments}

\appendix

\section{\label{app:DQD}Alternative examples of hierarchical environment}

\begin{figure}
\begin{centering}
\includegraphics[width=1\columnwidth]{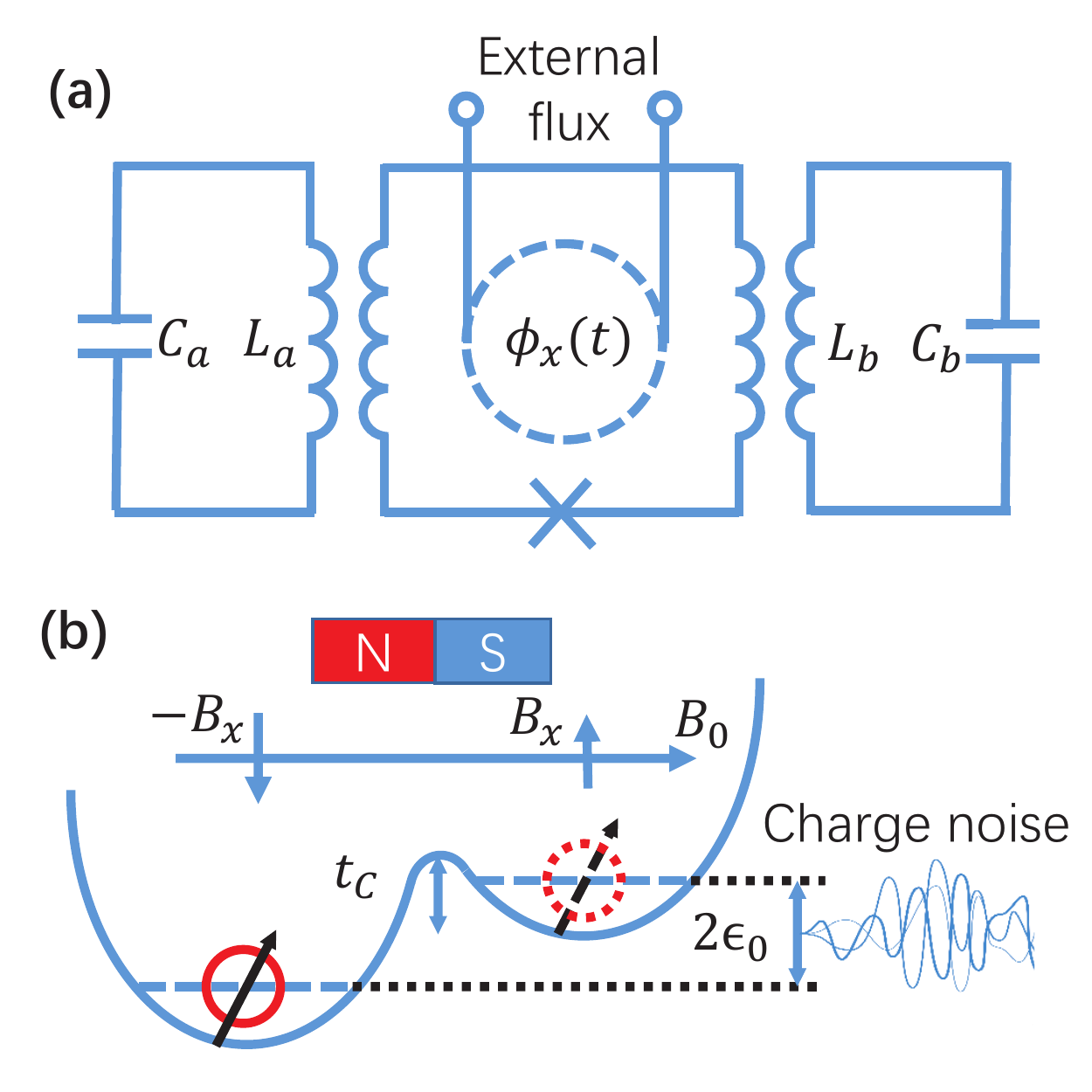}
\par\end{centering}
\caption{\label{fig:AltHE}Alternative examples of hierarchical environment.
(a) Another possible realization of J-C model Eq.~ in circuit-QED
systems \cite{You2005PT,You2011Nature}. Two micro-resonators are
coupled by a radio frequency SQUID coupler \cite{Tian2008NJP}. (b)
Single electron in double quantum dot. A nano-magnet produces the
gradient of the magnetic field to enhance the spin-orbit interaction.}
\end{figure}

In Fig.~\ref{fig:1}, we discuss a scenario that quantum system is
indirectly coupled to the bath through ``pseudo-environment'' illustrated
by J-C model coupled to an external bath. Besides the example in cavity-QED
system discussed in the main text, we also present an alternative
realization in circuit-QED system \cite{You2005PT,You2011Nature}.
As shown in Fig.~\ref{fig:AltHE}~(a), one of the resonator can
be defined as the quibt when the gap between the ground state and
the excited state is huge. Then, the other resonator can be modeled
as a harmonic oscillator. The tunable (time-dependent) coupling is
realized by the external flux $\phi_{x}(t)$, which is discussed in
Ref. \cite{Tian2008NJP}. Finally, the model can be described by the
Hamiltonian in Eq.~(\ref{eq:HJC}), and some other tunable coupling
schemes such like flux qubit coupled to a resonator are also discussed
in Refs. \cite{Brink2005NJP,Averin2003PRL,Plourde2004PRB,Hime2006Science}.

Besides the cavity-QED system and the circuit-QED system, a huge category
of physical models of hierarchical environment is often studied in
the semiconductor quantum dot system. In quantum dots, the spin degree
of freedom is rarely coupled to external noises resulting in their
super-long coherence time \cite{Burkard2020NRP}. Particularly, with
the help of isotopic purification that suppresses magnetic noise from
surrounding nuclear spins, the coherence time of spin qubit can exceed
one second \cite{Tyryshkin2012NM}, making it to be hopeful candidate
of the quantum processor. However, through the spin-orbit interaction,
spin states can indirectly coupled to noisy environments, which causes
spin relaxation/dephasing \cite{Zhao2016SR,Zhao2018SR}. Here, we
analyze a double quantum dot model which is used to realize spin-photon
interface \cite{Mi2018N}, where a single electron tunneling in double
quantum dots, forming two charge (orbital) states $|L\rangle$ and
$|R\rangle$. The applied magnetic field $B_{0}$ along $z$-direction
produces two spin states $|\uparrow\rangle$ and $|\downarrow\rangle$.
A nano-magnet produces a field gradient $B_{x}$ in $x$-direction,
making the magnetic field in left dot and right are different. The
schematic diagram is plotted in Fig.~\ref{fig:AltHE}~(b). In the
basis $\{|L,\uparrow\rangle,|L,\downarrow\rangle,|R,\uparrow\rangle,|R,\downarrow\rangle\}$,
the Hamiltonian can be written as 
\begin{equation}
H_{DQD}=\epsilon_{0}\tau_{z}+t_{C}\tau_{x}+B_{0}\sigma_{z}+B_{x}\sigma_{x}\tau_{x},
\end{equation}
where $\tau_{i}$ and $\sigma_{i}$ are the Pauli matrices in orbital
space and spin space respectively, $\epsilon_{0}$ is the detuning
between left and right dot and $t_{C}$ is the tunneling barrier.

In this example the charge (orbital) degrees of freedom plays the
role of ``pseudo-environment'' $H_{E}$ shown in Fig.~\ref{fig:1},
connecting electron spin to the ``bath'' such as charge fluctuation
or phonon noise. Practically, a fluctuation in magnetic field $B_{0}\rightarrow B_{0}+\delta B(t)$
causing a dephasing, similar to the case discussed in Eq.~(\ref{eq:wt}).
Alternatively, a charge noise in the detuning $\epsilon_{0}\rightarrow\epsilon_{0}+\delta\epsilon(t)$
can drive the electron continuously jumping between two dots thus
suffering different $x$-direction magnetic field. This is quite similar
to the example we discussed in Sec.~\ref{sec:3}, which is a time-dependent
interaction between the system and the pseudo-environment.

Some other examples can be also found in recent the experimental progress
\cite{Burkard2020NRP}. For example, when the semiconductor quantum
dots are coupled to a cavity, the system can be just roughly described
by a simplified J-C model \cite{Burkard2020NRP}. The mathematical
description of the dynamics of the system would be quite similar to
the results presented in this paper.

Anyway, in semiconductor quantum dots, the electron spin is typically
indirectly coupled to the environment through spin-orbit interaction,
which naturally provides a huge category of examples of hierarchical
environment. It is worth to note that in quantum dot systems, the
charge noise mainly plays the role of the classical noise. For example,
the fluctuation of the magnetic field often causes the dephasing of
the qubit. In this case, the classical noise is in the form of $\eta(t)$
as shown in Eq.~(\ref{eq:wt}).

\noindent 

\section{\label{app:MEQ}Derivation of master equation}

\subsection{noise in $\eta(t)$}

In this section, we show how to derive the master equation from the
NMQSD equation (\ref{eq:QSDO}). Taking the $\eta(t)$ noise as an
example and assuming the coupling $G(t)=G_{0}\sin(kx_{0})$ is a constant,
the time derivative of the density matrix can be written as
\begin{eqnarray}
\frac{d}{dt}\rho_{ac} & = & \frac{d}{dt}\left\langle M\left\{ |\psi_{t}\rangle\langle\psi_{t}|\right\} \right\rangle ,\nonumber \\
 & = & \left\langle M\left\{ \frac{d}{dt}|\psi_{t}\rangle\langle\psi_{t}|+|\psi_{t}\rangle\frac{d}{dt}\langle\psi_{t}|\right\} \right\rangle ,\nonumber \\
 & = & \langle M\{[-iH_{{\rm 0}}-i\frac{\eta(t)}{2}\sigma_{z}+az_{t}^{\ast}-a^{\dagger}\bar{O}]|\psi_{t}\rangle\langle\psi_{t}|\nonumber \\
 & + & |\psi_{t}\rangle\langle\psi_{t}|[iH_{{\rm 0}}+i\frac{\eta(t)}{2}\sigma_{z}+a^{\dagger}z_{t}-\bar{O}^{\dagger}a]\}\rangle,\label{eq:drho}
\end{eqnarray}
where $H_{0}=\frac{\omega_{0}}{2}\sigma_{z}+\Omega a^{\dagger}a+G_{0}\sin(kx_{0})(a\sigma_{+}+a^{\dagger}\sigma_{-})$,
and the time derivative of the stochastic state vector $\frac{d}{dt}|\psi_{t}\rangle$
is substituted by the NMQSD Eq.~(\ref{eq:QSDO}). Noticing that the
statistical mean $M_{1}\{\cdot\}$ and $\langle\cdot\rangle$ are
averaging over noises, if the kernel functions are independent of
noise variables, the mean values will be the kernel functions themselves.

In order to compute the right-hand-side of Eq.~(\ref{eq:drho}),
we introduce two lemmas and prove them by using the Novikov theorem
and the chain rule of the functional derivative.

Lemma 1.1

\begin{align}
 & \left\langle M\left\{ -i\eta(t)|\psi_{t}\rangle\langle\psi_{t}|\right\} \right\rangle \nonumber \\
 & =\left\langle M\left\{ |\psi_{t}\rangle\langle\psi_{t}|\bar{D}^{\dagger}-\bar{D}|\psi_{t}\rangle\langle\psi_{t}|\right\} \right\rangle ,\label{eq:lemma1.1}
\end{align}
where $\bar{D}=i\int_{0}^{t}ds\left\langle \eta(t)\eta(s)\right\rangle D(t,s,\eta)=i\int_{0}^{t}ds\alpha_{3}(t,s)\frac{\delta}{\delta\xi(s)}$
is a functional derivative operator.

Proof: According to the chain rule for functional derivative the term
$\left\langle M\left\{ -i\eta(t)|\psi_{t}\rangle\langle\psi_{t}|\right\} \right\rangle $
can be expanded as
\begin{align}
 & \left\langle M\left\{ -i\eta(t)|\psi_{t}\rangle\langle\psi_{t}|\right\} \right\rangle =-iM\left\{ \left\langle \eta(t)|\psi_{t}\rangle\langle\psi_{t}|\right\rangle \right\} ,\nonumber \\
 & =-iM\left\{ \int_{0}^{t}ds\left\langle \eta(t)\eta(s)\right\rangle \left\langle \frac{\delta(|\psi_{t}\rangle\langle\psi_{t}|)}{\delta\eta(s)}\right\rangle \right\} ,\nonumber \\
 & =-iM\left\{ \int_{0}^{t}ds\left\langle \eta(t)\eta(s)\right\rangle \left\langle \frac{\delta|\psi_{t}\rangle}{\delta\eta(s)}\langle\psi_{t}|+|\psi_{t}\rangle\frac{\delta\langle\psi_{t}|}{\delta\eta(s)}\right\rangle \right\} ,\nonumber \\
 & =\left\langle M\left\{ |\psi_{t}\rangle\langle\psi_{t}|\bar{D}^{\dagger}-\bar{D}|\psi_{t}\rangle\langle\psi_{t}|\right\} \right\rangle ,
\end{align}
where the functional derivative $\frac{\delta}{\delta\eta(s)}$ is
replaced by a time-dependent (maybe also noise dependent) operator
$D(t,s,\eta)$ defined as
\begin{equation}
\frac{\delta}{\delta\eta(s)}|\psi_{t}\rangle=D(t,s,\eta)|\psi_{t}\rangle.\label{eq:D-1}
\end{equation}
Since the stochastic state vector can be obtained by a stochastic
evolution operator acting on the initial state $|\psi_{t}\rangle=U(t,\eta)|\psi_{0}\rangle$,
then the existence of such an operator $D$ can be proved as
\begin{align}
\frac{\delta}{\delta\eta(s)}|\psi_{t}\rangle & =\left[\frac{\delta}{\delta\eta(s)}U(t,\eta)\right]|\psi(0)\rangle,\nonumber \\
 & =\left[\frac{\delta}{\delta\eta(s)}U(t,\eta)\right]U^{-1}(t,\eta)|\psi_{t}\rangle.
\end{align}
As a result, the operator $D$ can be formally defined as $D(t,s,\eta)=\left[\frac{\delta}{\delta\eta(s)}U(t,\eta)\right]U^{-1}(t,\eta)$.
According to Eq.~(\ref{eq:D-1}) and noticing that the order of the
averaging operation $M_{1}\{\cdot\}$ and $\langle\cdot\rangle$ can
be swapped, lemma 1.1 is proved.

Lemma 1.2

\begin{equation}
\left\langle M_{1}\left\{ z_{t}^{*}|\psi_{t}\rangle\langle\psi_{t}|\right\} \right\rangle =\left\langle M_{1}\left\{ \bar{O}|\psi_{t}\rangle\langle\psi_{t}|\right\} \right\rangle .
\end{equation}
The proof of this lemma for the quantized noise is different from
classical noise. It can be also found in Refs.~\cite{Yu1999PRA}.

With lemma 1.1 and 1.2, one can substitute the results into Eq.~(\ref{eq:drho})
to obtain a master equation as
\begin{align}
 & \frac{d}{dt}\rho=-i\left[H_{0},\rho\right]\nonumber \\
 & +\left[a,\left\langle M\left\{ |\psi_{t}\rangle\langle\psi_{t}|\bar{O}^{\dagger}\right\} \right\rangle \right]+\left[\left\langle M\left\{ \bar{O}|\psi_{t}\rangle\langle\psi_{t}|\right\} \right\rangle ,a^{\dagger}\right]\nonumber \\
 & +\left[\sigma_{z},\left\langle M\left\{ |\psi_{t}\rangle\langle\psi_{t}|\bar{D}^{\dagger}\right\} \right\rangle \right]+\left[\left\langle M\left\{ \bar{D}|\psi_{t}\rangle\langle\psi_{t}|\right\} \right\rangle ,\sigma_{z}\right].\label{eq:MEQraw}
\end{align}
The operator $\bar{D}$ can be determined by the consistency condition
$\frac{\delta}{\delta\eta(s)}\frac{d}{dt}|\psi_{t}\rangle=\frac{d}{dt}\frac{\delta}{\delta\eta(s)}|\psi_{t}\rangle$.
The left-hand-side can be written as
\begin{align}
\frac{\delta}{\delta\eta(s)}\frac{d}{dt}|\psi_{t}\rangle & =\frac{\delta}{\delta\eta(s)}\left(H_{{\rm eff}}|\psi_{t}\rangle\right)\nonumber \\
 & =\frac{\delta}{\delta\eta(s)}H_{{\rm eff}}|\psi_{t}\rangle+H_{{\rm eff}}\frac{\delta}{\delta\eta(s)}|\psi_{t}\rangle\nonumber \\
 & =\frac{\delta}{\delta\eta(s)}H_{{\rm eff}}|\psi_{t}\rangle+H_{{\rm eff}}D|\psi_{t}\rangle,
\end{align}
while the right-hand-side can be written as
\begin{align}
\frac{d}{dt}\frac{\delta}{\delta\eta(s)}|\psi_{t}\rangle & =\frac{d}{dt}\left[D(t,s,\eta)|\psi_{t}\rangle\right]\nonumber \\
 & =\frac{d}{dt}D|\psi_{t}\rangle+D\frac{d}{dt}|\psi_{t}\rangle\nonumber \\
 & =\frac{d}{dt}D|\psi_{t}\rangle+DH_{{\rm eff}}|\psi_{t}\rangle.
\end{align}
Equating left-hand-side and right-hand-side, one can obtain
\begin{equation}
\frac{d}{dt}D=\left[H_{{\rm eff}},D\right]+\frac{\delta}{\delta\eta(s)}H_{{\rm eff}}.\label{eq:dD}
\end{equation}
Integrating over an infinitesimal time interval $s_{-}<t<s_{+}$ around
$t=s$, 
\begin{align}
 & D(t=s_{+},s,\eta)-D(t=s_{+},s,\eta)\nonumber \\
 & =\int_{s_{-}}^{s_{+}}dt\left\{ -i\frac{1}{2}\delta(t,s)\sigma_{z}+\left[H_{{\rm eff}},D\right]\right\} ,
\end{align}
since the functional derivative $\frac{\delta}{\delta\eta(s)}$ is
only non-zero on the term $\frac{\eta(t)}{2}\sigma_{z}$ in $H_{{\rm eff}}$.
In the integral, the contribution from the second term is zero since
$\left[H_{{\rm eff}},D\right]$ is finite and the integration interval
is close to zero, $s_{+}-s_{-}\rightarrow0$. Therefore, at the boundary
$s=t$, the boundary condition reads $D(t=s,s,\eta)=-i\frac{1}{2}\sigma_{z}$.

Similarly, one can also use the consistency condition $\frac{\delta}{\delta z_{t}^{*}}\frac{d}{dt}|\psi_{t}\rangle=\frac{d}{dt}\frac{\delta}{\delta z_{t}^{*}}|\psi_{t}\rangle$
to obtain
\begin{equation}
\frac{d}{dt}O=\left[H_{{\rm eff}},D\right]+a^{\dagger}\frac{\delta}{\delta z_{s}^{*}}\bar{O},\label{eq:dO}
\end{equation}
with the boundary condition $O(t=s,s,z^{*})=a$.

In most cases, we can approximately assume the operators $D$ and
$O$ are all noise-independent \cite{Xu2014JPA}. Then, the master
equation can be further simplified as
\begin{equation}
\frac{d}{dt}\rho=-i\left[H_{{\rm 0}},\rho\right]+\left\{ \left[a,\rho\bar{O}^{\dagger}\right]+\left[\sigma_{z},\rho\bar{D}^{\dagger}\right]+{\rm h.c.}\right\} ,
\end{equation}
Here, the master equation does not formally contain a cross-term combining
the impact from both $z_{t}^{*}$ and $\eta(t)$, this is because
the two noises are independent (not correlated noise). More details
on the discussion of correlated noise can be found in \cite{Corn2009QIP,Jing2015PRA}.
However, if we compute the exact solution of operators $D$ and $O$,
one may find that $O$ has an impact on $D$ and $D$ also has an
impact on $O$. Therefore, interference effect between two noises
may exist. One example can be found in Ref.~\cite{Zhao2017AoP}.

\subsection{noise in $\xi(t)$}

In the above subsection, we have shown the detailed derivation of
the master equation in the presence of $\eta(t)$ noise, now, we will
briefly show the case of $\xi(t)$ noise discuss the difference between
them. When, $\eta(t)=0$ and $\xi(t)\neq0$, the time derivative of
$|\psi_{t}\rangle$ is 
\begin{eqnarray}
\frac{d}{dt}\rho & = & \langle M\{[-iH_{{\rm 0}}^{\prime}-i\frac{G(t)}{2}V+az_{t}^{\ast}-a^{\dagger}\bar{O}]|\psi_{t}\rangle\langle\psi_{t}|\nonumber \\
 & + & |\psi_{t}\rangle\langle\psi_{t}|[iH_{{\rm 0}}^{\prime}+i\frac{G(t)}{2}V+a^{\dagger}z_{t}-\bar{O}^{\dagger}a]\}\rangle,\label{eq:drho-1}
\end{eqnarray}
where $H_{0}^{\prime}=\frac{\omega_{0}}{2}\sigma_{z}+\Omega a^{\dagger}a$,
$V=a\sigma_{+}+a^{\dagger}\sigma_{-}$. Now, the noise is included
in the coupling $G(t)$. Similar to Eq.~(\ref{eq:lemma1.1}), one
can prove

\begin{align}
 & \left\langle M\left\{ -iG(t)|\psi_{t}\rangle\langle\psi_{t}|\right\} \right\rangle \nonumber \\
 & =\left\langle M\left\{ |\psi_{t}\rangle\langle\psi_{t}|\bar{D}_{G}^{\dagger}-\bar{D}_{G}|\psi_{t}\rangle\langle\psi_{t}|\right\} \right\rangle ,
\end{align}
where $\bar{D}=i\int_{0}^{t}ds\left\langle G(t)G(s)\right\rangle D_{G}(t,s,G)=i\int_{0}^{t}ds\left\langle G(t)G(s)\right\rangle \frac{\delta}{\delta G(s)}$
is a functional derivative operator with the boundary condition $D_{G}(t,s=t,G)=V$.
Because we take the functional derivative with respect to $G(t)$
here, the correlation is in the form of $\left\langle G(t)G(s)\right\rangle $.
However, if $\langle\xi(t)\xi(s)\rangle=\alpha_{2}(t,s)$ is known,
it is straightforward to compute $\left\langle G(t)G(s)\right\rangle $.
Eventually, the master equation can be written as

\begin{equation}
\frac{d}{dt}\rho=-i\left[H_{{\rm 0}}^{\prime},\rho\right]+\left\{ \left[a,\rho\bar{O}^{\dagger}\right]+\left[V,\rho\bar{D}_{G}^{\dagger}\right]+{\rm h.c.}\right\} .
\end{equation}

\subsection{\label{subApp:O}Equation for operator $O$}

\begin{figure}
\begin{centering}
\includegraphics[width=1\columnwidth]{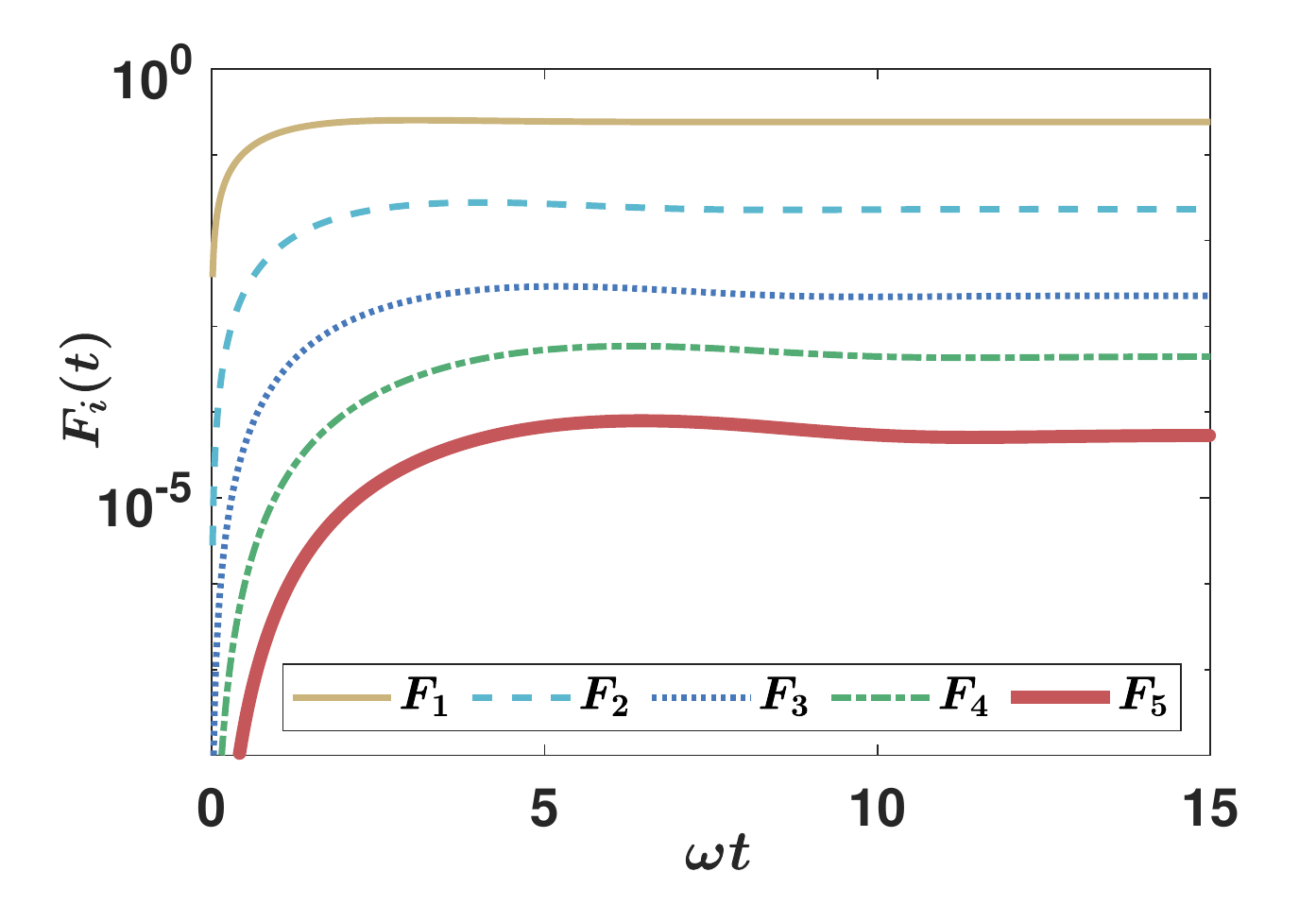}
\par\end{centering}
\caption{\label{fig:Fi}Time evolution of the coefficients in operator $\bar{O}$.
The parameters are $\omega_{0}=\Omega=1,\Gamma_{1}=1$, and $\gamma_{1}=0.5$.}
\end{figure}

According to the consistency condition (\ref{eq:dO}), we find the
solution of $O$ operator in Eq.~(\ref{eq:MEQhalf}) is 
\begin{equation}
O(t,s,z^{*})=\sum_{i=1}^{4}f_{i}(t,s)O_{i}+\int_{0}^{t}f_{5}(t,s,s^{\prime})z_{s^{\prime}}^{*}ds^{\prime}O_{5},\label{O}
\end{equation}
with the basis operators $O_{1}=a$, $O_{2}=\sigma_{-}aa^{\dagger}$,
$O_{3}=\sigma_{-}a^{\dagger}a$, $O_{4}=\sigma_{z}a$, and $O_{5}=\sigma_{-}a$.
The coefficients satisfy the following equations 
\begin{widetext}
\begin{eqnarray}
\partial_{t}f_{1}(t,s) & = & i\Omega f_{1}+i\frac{G}{2}f_{2}-i\frac{G}{2}f_{3}-f_{1}F_{1}-f_{4}F_{4},\\
\partial_{t}f_{2}(t,s) & = & i\omega f_{2}+iGf_{1}-iGf_{4}-f_{1}F_{2}+f_{4}F_{2},\\
\partial_{t}f_{3}(t,s) & = & i\omega f_{3}-iGf_{1}-iGf_{4}+f_{1}F_{2}+f_{4}F_{2}-2f_{3}F_{4}-F_{5}^{\prime}(t,s),\label{pf3}\\
\partial_{t}f_{4}(t,s) & = & i\Omega f_{4}-i\frac{G}{2}f_{2}-i\frac{G}{2}f_{3}-f_{4}F_{1}-f_{1}F_{4},\\
\partial_{t}f_{5}(t,s,s^{\prime}) & = & i\omega f_{5}+i\Omega f_{5}-f_{5}F_{1}-f_{5}F_{4}-f_{1}F_{5}^{\prime}+f_{4}F_{5}^{\prime},
\end{eqnarray}
\end{widetext}

where $F_{i}(t)=\int_{0}^{t}\alpha(t,s)f_{i}(t,s)ds$ ($i=1,2,3,4$)
and $F_{5}^{\prime}(t,s^{\prime})=\int_{0}^{t}\alpha(t,s)f_{5}(t,s,s^{\prime})ds$,
and the initial conditions are 
\begin{equation}
f_{1}(t,t)=1,\quad f_{i}(t,t)=0,\quad(i=2,3,4)
\end{equation}
\begin{equation}
f_{5}(t,t,s^{\prime})=0,\quad f_{5}(t,s,t)=f_{3}-f_{2}.
\end{equation}
As we have discussed in Sec.~\ref{sec:2Model}, both Eqs.~(\ref{eq:MEQG})
and (\ref{eq:MEQsz}) are reduced to Markovian form when $\alpha_{1}(t,s)=\Gamma_{1}\delta(t,s)$
leading to a simplified $\bar{O}$ with only the $O_{1}$ term, because
$F_{1}(t)=\int_{0}^{t}\Gamma_{1}\delta(t,s)f_{1}(t,s)ds=\frac{\Gamma_{1}}{2}$
and all the other coefficients $F_{i}=0$ ($i=2,3,4,5$), where $F_{5}=\int_{0}^{t}\alpha(t,s)F_{5}^{\prime}(t,s)ds$.
Therefore, the other terms can be regarded as the non-Markovian corrections
to the $O_{1}$ term. Typically these correction terms are quite small,
particularly the noise-dependent term $O_{5}$ is much smaller than
the other terms (Detailed study about the impact of noise-dependent
term can be found in Ref. \cite{Xu2014JPA} with a two-qubit dissipative
model. Here we simply use numerical simulation to show $\frac{|F_{5}|}{|F_{1}|}\approx10^{-4}$
in Fig.~\ref{fig:Fi}. Thus, it is safe to neglect the $O_{5}$ term
and Eq. (\ref{O}) is reduced to 
\begin{equation}
O(t,s)=\sum_{i=1}^{4}f_{i}(t,s)O_{i},\label{eq:O2}
\end{equation}
where $O(t,s)$ becomes independent of $z_{t}^{*}$. Besides, it is
more important to note that the memory time used to generate Fig.~\ref{fig:Fi}
is $\gamma_{1}=0.5$, corresponding to a relative non-Markovian value.
When we have a shorter memory time $1/\gamma_{1}$ ($\gamma_{1}$
is larger), the correlation function will even close to a delta-function.
Then, the delta-like correlation function leads to $F_{5}\rightarrow0$
as we have discussed. In that case, it is even more reasonable to
neglect $F_{5}$ in Eq.~(\ref{eq:O2}).

In the most general case, there is a systematic way to expand $O$
operator up to certain orders \cite{Yu1999PRA,Li2014PRA}. With a
noise dependent $O$ operator, the master equation (\ref{eq:MEQraw})
can be also simplified to a Lindblad form by following the procedure
developed in Ref.~\cite{Chen2014PRA}. In the derivation of the master
equation, we also assume the temperature of the bath is zero. The
master equation of the finite temperature case can be obtained by
using the technique in Ref.~\cite{Yu2004PRA}. Several examples can
be found in Refs.~\cite{Zhao2009OC,Zhao2017AoP,Zhao2019OEa,Shi2013PRA}.

\noindent 

\section{\label{app:OUnoise}Arbitrary correlation function}

In the main text, we choose the O-U correlation function in Eq.~(\ref{eq:OUN1}),
which, by definition, should be the correlation in Eq.~(\ref{QSD})
$\alpha_{1}(t,s)=\sum_{i}|g_{i}|^{2}e^{-i\omega_{i}(t-s)}$. We would
like to emphasize that the NMQSD equation as well as the master equations
we derived do not depend on the choice of correlation functions. The
dynamical equations are applicable to all types of correlation functions.
The choice of O-U correlation function is based on the need of showing
transition from Markovian to non-Markovian regime and the impact of
the memory time of the bath. Here, we show an example that the widely
used $1/f$ noise can be obtained by summation of Lorentzian spectrum
density in Eq.~(\ref{eq:gw}). Consider a set of Lorentzian spectrum
density with different memory time $\gamma_{1}$ and statistical weight
$W(\gamma_{1})=\frac{1}{\gamma_{1}^{2}}d\gamma_{1}$, then the summation
of these Lorentzian spectrum leads to 
\begin{eqnarray}
\tilde{g}(\omega) & = & \int_{\gamma_{1L}}^{\gamma_{1H}}\frac{1}{\gamma_{1}^{2}}g(\omega)d\gamma_{1}\nonumber \\
 & = & \frac{\Gamma_{1}}{2\pi}\frac{1}{\omega}\arctan\left(\frac{\gamma_{1}}{\omega}\right)|_{\gamma_{1L}/\omega}^{\gamma_{1H}/\omega}\approx\frac{\Gamma_{1}}{4}\frac{1}{\omega},
\end{eqnarray}
which is proportional to $1/\omega$, thus called $1/f$ noise. The
cut-off frequencies are labeled as $\gamma_{1H}$ and $\gamma_{1L}$.

In the numerical simulation, a noise $z(t)$ satisfying an arbitrary
correlation function $M[z(t)z(s)^{*}]=\alpha(t-s)$ can be generated
by two independent real Gaussian noises $z_{1}(t)$ and $z_{2}(t)$
satisfying $M[z_{1}(t)]=M[z_{2}(t)]=0$, and $M[z_{1}(t)z_{1}(s)]=M[z_{2}(t)z_{2}(s)]=\delta(t-s)$.
One can first transform the correlation function into frequency domain
as $|G(\omega)|^{2}=\int_{-\infty}^{\infty}\frac{1}{2\pi}\alpha(\tau)e^{-i\omega\tau}d\tau$,
then taking the inverse Fourier transformation of the square root
$G(\omega)=\sqrt{|G(\omega)|^{2}}$ as $R(t)=\int_{-\infty}^{\infty}\frac{1}{2\pi}G(\omega)e^{i\omega t}d\omega$.
Finally the noise $z(t)$ can be constructed by 
\begin{equation}
z(t)=\int_{-\infty}^{\infty}dsR(s)\frac{z_{1}(t-s)+iz_{2}(t-s)}{\sqrt{2}}.
\end{equation}

\begin{figure}[H]
\begin{centering}
\includegraphics[width=3.5in]{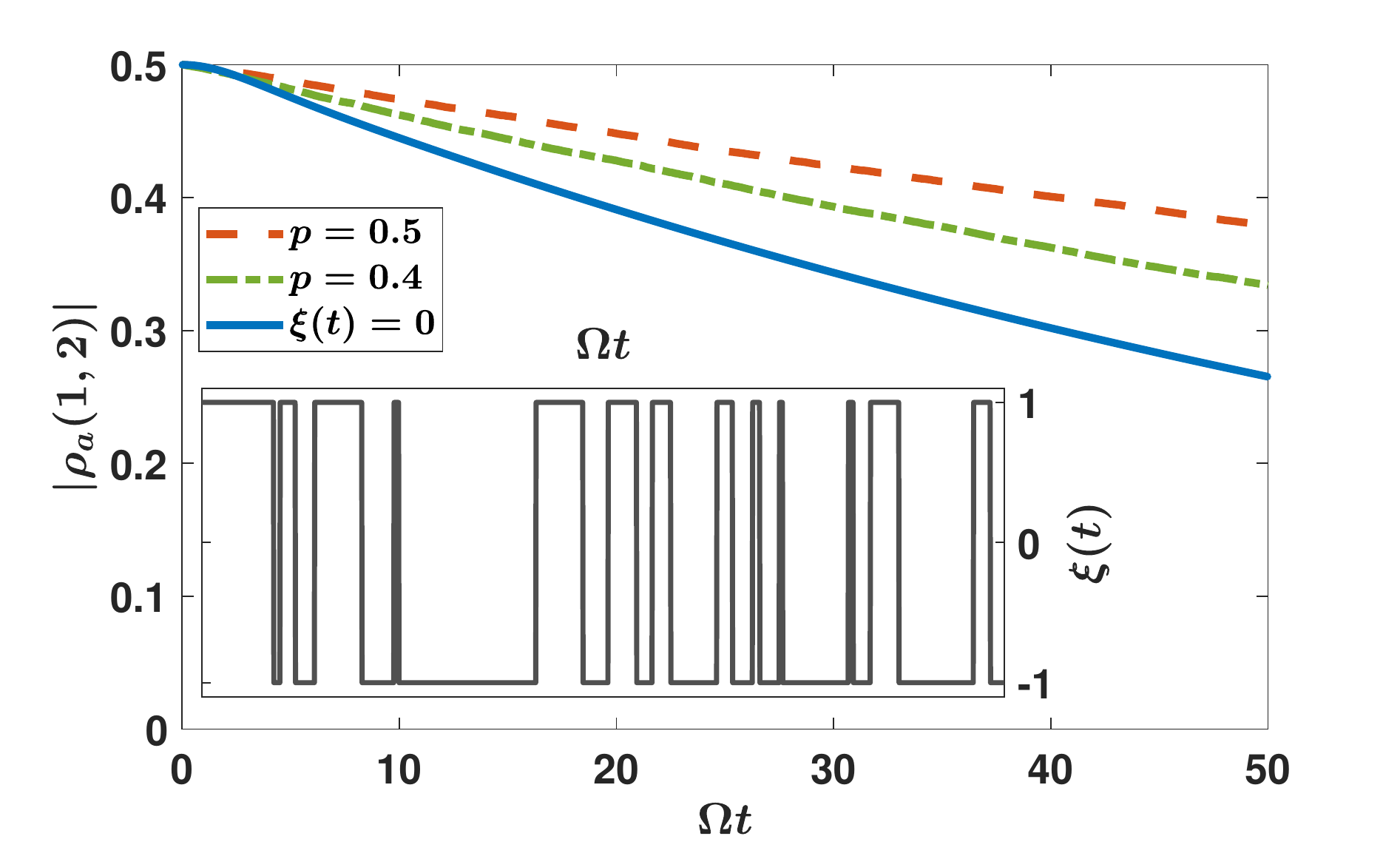} 
\par\end{centering}
\caption{\label{TeleNoise} Decoherence process of the atom with different
$p$. The inset-plot is a schematic diagram of a single random realization
of the telegraph noise $\xi(t)$.}
\end{figure}

Here, we also provide an example of applying other types of noises.
According to Ref. \cite{Cheng2008PRA}, the so-called telegraph noise
is generated as follows. The total evolution time period is separated
into $N$ small intervals $\{t_{0},t_{1},t_{2}\ldots t_{N}\}$, in
each interval $(t_{i}<t<t_{i+1})$, $x(t)=\pm1$, as shown in Fig.
\ref{TeleNoise} (inset-plot). At time $t_{i}$ the variable $\xi(t)$
may change its sign as $\xi(t_{i})=-\xi(t_{i-1})$ with probability
$p$, and according to the normalization, the possibility that $\xi(t)$
keeps its value unchanged is $(1-p)$. Fig.~\ref{TeleNoise} shows
how does the jumping probability $p$ affect the decoherence. When
the random variable $\xi(t)$ changes its sign more frequently (larger
$p$), the coherence protection is better. According to the discussion
in Sec.~\ref{subsec:Mechanism}, a larger $p$ implies a choppier
noise, thus results in a better protection. This is reflected in Fig.
\ref{TeleNoise}, small $p$ typically results in a bad coherence
protection.

\noindent 

\section{\label{app:outlook}Coherence and entanglement}

In Fig.~\ref{fig:CohNeg}, we use the negativity \cite{Vidal2002PRA}
$N[\rho_{ac}(t)]$ to quantify the entanglement instead of $C(\rho_{ac})$
in Eq.~(\ref{eq:Crho}). This is because it is relatively easy to
derive the analytical expression of $C(\rho_{ac})$, but it is only
applicable to $2\times2$ system. In the numerical analysis, we are
aiming at characterizing the general case in which the cavity may
have more than one photon number. Thus, we choose $N(\rho_{ac})$
which is applicable to $2\times3$ or even $2\times N$ cases. However,
both $C(\rho_{ac})$ and $N(\rho_{ac})$ are entanglement monotones
\cite{Vidal2000JoMO}, so that they have one-to-one correspondence
and describe the same physical picture in this model.

The numerical results in Fig.~\ref{fig:CohNeg}~(a) show that a
stronger noise $\xi(t)$ (reflected by a larger $\Gamma_{2}$) leads
to a better performance of the coherence protection. To illustrate
our hypothesis that the coherence is protected because the noise can
weaken the entanglement generation between the atom and its environment,
the time evolution of atom-cavity entanglement $N(t)$ is also plotted
with the same parameters in Fig.~\ref{fig:CohNeg}~(b). The coherence
and the entanglement evolution almost follow the identical pattern,
which implies the atomic coherence and atom-cavity entanglement are
highly correlated. Here, in order to present this correlation clearly,
we intentionally plot the quantity ``$-N(t)$'', which is always
negative and a smaller value corresponds to a stronger entanglement.

It is worth to note that the system plus environment is actually a
tripartite system consists the atom (S), the cavity (E), and the bath
(B). The results shown in Fig.~\ref{fig:CohNeg} only proves the
correlation between the atomic coherence and the atom-cavity entanglement.
From the view of the atom, both the cavity and the bath are its ``environments''.
Since the total state (living in the Hilbert space of S-E-B) is always
a pure state, the Von Neumann entropy \cite{Vedral1997PRL} defined
based on the reduced density matrix of the atom $\rho_{a}$ is also
a measure of the entanglement between the atom and its ``environment''.
The coherence loss itself indicates the atom is already entangled
with its ``environment''. Theoretically, both atom-cavity (S-E)
entanglement and atom-bath (S-B) entanglement can lead to atomic coherence
loss. Or, the coherence is not related to entanglement \cite{Perez2020PRRes,Ai2010PRA,Linke2018PRA}.
Here, we highlight the atom-cavity entanglement presented in Fig.~\ref{fig:CohNeg}
to emphasize the coherence can be protected by just cutting off the
entanglement generation between the atom and cavity (S-E). This is
because only the cavity (E) is directly coupled to the atom (S) and
the effective coupling to the external bath is indirect (through the
cavity). Eventually, the atom will inevitably decay to ground state
and loss the coherence for this dissipative model (in long-time limit,
dissipation to the bath will be the main reason for atomic coherence
loss), but cutting off the atom-cavity entanglement generation at
the early stage can dramatically slow down this process because the
bath can be only coupled to the atom through the cavity.

\noindent \bibliographystyle{apsrev}
\bibliography{NIDS2}

\end{document}